\documentclass[11pt,a4paper]{article}
\pdfoutput=1 % if your are submitting a pdflatex (i.e. if you have
\usepackage{jheppub}
\usepackage{latexsym}
\usepackage{revsymb}

\newcommand{\Bo}{{}^{8}\mathrm{B}}
\newcommand{\Be}{{}^{7}\mathrm{Be}}
\newcommand{\Kr}{{}^{85}\mathrm{Kr}}
\newcommand{\Rb}[1]{{}^{#1}\mathrm{Rb}}
\newcommand{\He}[1]{{}^{#1}\mathrm{He}}
\newcommand{\Bi}[1]{{}^{#1}\mathrm{Bi}}
\newcommand{\Rn}[1]{{}^{#1}\mathrm{Rn}}
\newcommand{\Po}[1]{{}^{#1}\mathrm{Po}}
\newcommand{\Pb}[1]{{}^{#1}\mathrm{Pb}}
\newcommand{\C}[1]{{}^{#1}\mathrm{C}}
\newcommand{\Li}[1]{{}^{#1}\mathrm{Li}}

\preprint{EURONU-WP6-12-52, IFIC/12-49}

\title{Constraining Non-Standard Interactions of the Neutrino with Borexino}

\author[a]{Sanjib Kumar Agarwalla,}
\author[b]{Francesco Lombardi,}
\author[c]{and Tatsu Takeuchi$\,$}

\affiliation[a]{Instituto de F\'{\i}sica Corpuscular, CSIC-Universitat de Val\`encia, Apartado de Correos 22085, E-46071 Valencia, Spain}
\affiliation[b]{Universit\`a degli Studi dell'Aquila, Dipartimento di Fisica, L'Aquila, Italy and\\
INFN, Laboratori Nazionali del Gran Sasso, Assergi (AQ), Italy}  
\affiliation[c]{Center for Neutrino Physics, Physics Department, Virginia Tech, Blacksburg, 24061 VA, USA}

\emailAdd{Sanjib.Agarwalla@ific.uv.es}
\emailAdd{francesco.lombardi@lngs.infn.it}
\emailAdd{takeuchi@vt.edu}

\abstract{
We use the Borexino 153.6 $\mathrm{ton\!\cdot\!year}$ data to place constraints on
non-standard neutrino-electron interactions, taking into account the uncertainties in the
$\Be$ solar neutrino flux and the mixing angle $\theta_{23}$, 
and backgrounds due to $\Kr$ and $\Bi{210}$ $\beta$-decay.
We find that the bounds are comparable to existing bounds from all other experiments.
Further improvement can be expected in Phase~II of Borexino due to the
reduction in the $\Kr$ background.
}

\arxivnumber{1207.3492}

\dedicated{We dedicate this paper to the memory of Raju Raghavan\\
who played a significant role in the Borexino experiment and\\
was the initiator and driving force behind this work.}

\begin{document}
\maketitle
\flushbottom
%%%%%%%%%%%%%%%%%%%%%%%%%%%%%%%%%%%%%%%%%%%%%%%%%%%%%%%%%%%%%%%%%%%%%%%%%%%%%%
%%%%%%%%%%%%%%%%%%%%%%%%%%%%%%%%%%%%%%%%%%%%%%%%%%%%%%%%%%%%%%%%%%%%%%%%%%%%%%
\section{Introduction}
\label{sec:intro}
%%%%%%%%%%%%%%%%%%%%%%

Various extensions of the Standard Model (SM), such as left-right symmetric models
and supersymmetric models with R-parity violation,
predict non-standard interactions (NSIs) of the neutrinos with
other fermions
\cite{Roulet:1991sm,Guzzo:1991hi,Barger:1991ae,Bergmann:1998ft,Bergmann:1999pk,Antusch:2008tz,Gavela:2008ra,Malinsky:2008qn,Ohlsson:2009vk,Medina:2011jh,Berezhiani:2001rs}.
The NSIs in those models are generated via the exchange of new massive particles and at low-energies
can be described by effective four fermion interactions of the form
\begin{equation}
\mathcal{L}_\mathrm{NSI} \;=\; -2\sqrt{2}\,G_{F}\,\varepsilon_{\alpha \beta}^{ff'C}
\bigl(
\overline{{\nu}_{\alpha}} \gamma^{\mu} P_L \nu_{\beta}
\bigr)
\bigl(
\overline{f} \gamma_{\mu} P_C f'
\bigr)
\;,
\label{NSI}
\end{equation}
where $G_F$ is the Fermi constant, 
$\alpha$ and $\beta$ are neutrino flavor indices,
$f$ and $f'$ label light SM fermions,
$C=\mbox{$L$ or $R$}$ is the chirality of the projection operator $P_C$, where $P_{L/R}=(1\mp\gamma_5)/2$,
and the dimensionless number $\varepsilon_{\alpha\beta}^{ff'C}$ parametrizes the strength of the interaction.

In a previous publication from 2002 \cite{Berezhiani:2001rt}, Raghavan,
together with Berezhiani and Rossi, discussed the potential of the Borexino
detector in placing constraints on the flavor-diagonal NSI parameters
\begin{equation}
\varepsilon_{\alpha L}\,\equiv\,\varepsilon_{\alpha\alpha}^{eeL}\;,\qquad
\varepsilon_{\alpha R}\,\equiv\,\varepsilon_{\alpha\alpha}^{eeR}\;,\qquad
\alpha\,=\,\mbox{$e$ or $\tau$}\;,
\label{diagonalNSI}
\end{equation}
via the measurement of the electron recoil spectrum in $\nu_\alpha e$ elastic scattering.
There, it was argued that due to the mono-energetic nature of the $\Be$ solar neutrinos,
Borexino would be able to place stronger constraints on $\varepsilon_{eR}$ and
$\varepsilon_{\tau R}$ than would be possible at Super-Kamiokande (SK) and
the Sudbury Neutrino Observatory (SNO) where the observed neutrinos are the $\Bo$ neutrinos 
with a continuous energy spectrum.\footnote{%
This possibility had been suggested earlier by Berezhiani and Rossi in Ref.~\cite{Berezhiani:1994hy}.}
The $\alpha=\mu$ case was not considered since the couplings
$\varepsilon_{\mu L}$ and $\varepsilon_{\mu R}$ were already constrained to the level
of $|\varepsilon_{\mu L/R}|< 0.03$ (at 90\% C.L.) \cite{Davidson:2003ha}
by the $\nu_\mu e$ scattering experiment CHARM~II \cite{Vilain:1994qy}.

Today, a decade later, with the Borexino experiment running smoothly and
having accumulated more than 153.6 $\mathrm{ton\!\cdot\!years}$ of data,
it is now possible to actually extract the constraints
on the flavor-diagonal NSI parameters discussed by Raghavan et al.
Constraints on the same parameters from various other experiments are also available
for comparison, some of which are quite new.
Bounds from the solar neutrino experiments (SK, SNO, etc.) and KamLAND 
can be found in Refs.~\cite{Davidson:2003ha}, \cite{Barranco:2005ps}, and \cite{Bolanos:2008km}.
Bounds from reactor and accelerator experiments have been compiled in
Ref.~\cite{Barranco:2007ej}, which includes the bounds from
$e^+e^- \rightarrow \nu\bar\nu\gamma$ measured at LEP \cite{ALEPH:2005ab},
$\nu_e e$ scattering measured at LSND \cite{Auerbach:2001wg},
and $\overline{\nu}_e e$ scattering measured at the reactor
experiments Irvine \cite{Reines:1976pv},
Rovno \cite{Derbin:1993wy}, and MUNU \cite{Daraktchieva:2003dr}. 
See also Refs.~\cite{Biggio:2009kv,Biggio:2009nt}.
Bounds have also been placed using
atmospheric neutrinos \cite{Friedland:2004ah} and MINOS \cite{Friedland:2006pi}. 
New bounds from the TEXONO reactor neutrino experiment can be found in Ref.~\cite{Deniz:2010mp}.

In this paper, we place constraints on the NSI parameters
$\varepsilon_{eL/R}$ and $\varepsilon_{\tau L/R}$ using
the Borexino 153.6 $\mathrm{ton\!\cdot\!year}$ data. 
We include in our analysis the uncertainties in the $\Be$ solar neutrino flux
and in the mixing angle $\theta_{23}$, 
and backgrounds from $\Kr$ and $\Bi{210}$ $\beta$-decay.
Taking these systematic uncertainties and backgrounds into account are particularly important
since they can mimic non-zero values of the parameters in question.
Indeed, as will be discussed in detail in the following, we find that 
unless the solar-neutrino flux-uncertainty
and the $\Kr$ $\beta$-decay background are reduced, 
increased statistics will not improve the bounds beyond 
what can be extracted from current Borexino data.
The parameters $\varepsilon_{\mu L/R}$ are not considered since the CHARM~II bound
still stands.

This paper is organized as follows.
In section~\ref{sec:nue}, we discuss how the electron recoil spectrum in 
$\nu_\alpha e$ scattering can be used to constrain the parameters $\varepsilon_{\alpha L/R}$,
$(\alpha=\mbox{$e$ or $\tau$})$
and how the above mentioned
systematic uncertainties and $\beta$-decay backgrounds can interfere with the extraction.
Section~\ref{sec:borexino} discusses how well systematic uncertainties and
backgrounds are understood in the Borexino experiment, 
section~\ref{sec:Solar_Model} discusses the uncertainty in the
$\Be$ neutrino flux in the Standard Solar Model,
and section~\ref{sec:angles} discusses the uncertainties in the neutrino oscillation probabilities which depend on our knowledge of the mixing angles $\theta_{12}$ and $\theta_{23}$.
In Section~\ref{sec:event-nsi} we look at the dependence of the electron recoil 
spectrum on the NSI parameters and backgrounds in detail, and how 
one can mimic the other.
Results of our fit are presented in section~\ref{sec:results},
and section~\ref{sec:summary} concludes.

%%%%%%%%%%%%%%%%%%%%%%%%%%%%%%%%%%%%%%%%%%%%%%%%%%%%%%%%%%%%%%%%%%%%%%%%%%%%%%
%%%%%%%%%%%%%%%%%%%%%%%%%%%%%%%%%%%%%%%%%%%%%%%%%%%%%%%%%%%%%%%%%%%%%%%%%%%%%%
\section{Neutrino-Electron Elastic Scattering}
\label{sec:nue}
%%%%%%%%%%%%%%%%%%%%%%

In the SM, the interaction between neutrino flavor $\alpha$ ($\alpha=e,\mu,\tau$) and the electron 
is described at low energies by the effective four fermion interaction
%
%%%%%%%%%%%%%%%%%%%%%%%%%%%%%%%%%%%%%
\begin{equation}
\mathcal{L}_\mathrm{SM} 
\;=\;
-2\sqrt{2}\,G_{F} (\bar{\nu}_{\alpha} \gamma^{\mu} P_L \nu_{\alpha})
\Bigl[\,
g_{\alpha L} (\bar{e} \gamma_{\mu} P_{L} e) + 
g_{\alpha R} (\bar{e} \gamma_{\mu} P_{R} e)
\,\Bigr].
\label{eqn1}
\end{equation}
%%%%%%%%%%%%%%%%%%%%%%%%%%%%%%%%%%%%%
%
The coupling constants at tree level are given by $g_{\alpha R} = \sin^2 \theta_{W}$ and  
$g_{\alpha L} = \sin^2\theta_{W} \pm \frac{1}{2}$, where the lower sign applies for 
$\alpha=\mu$ and $\tau$ (from $Z$ exchange only) and the upper sign applies for $\alpha=e$ (from both $Z$ and $W$ exchange).
Radiative corrections are small and can be ignored.\footnote{%
In the notation of Ref.~\cite{Erler:2012}, the coupling constants for the
$\alpha=\mu$ case, including radiative corrections,
are expressed as
\[
g_{\mu L}
\;=\;\dfrac{g_{V}^{\nu e}+g_{A}^{\nu e}}{2}
\;=\;\rho_{\nu e}\left(-\dfrac{1}{2}+\hat{\kappa}_{\nu e}\hat{s}^2_Z\right)\;,\qquad
g_{\mu R}
\;=\;\dfrac{g_{V}^{\nu e}-g_{A}^{\nu e}}{2}
\;=\;\rho_{\nu e}\left(-\dfrac{1}{2}\right)\;,
\]
where $\hat{s}^2_Z$ is the $\overline{\mathrm{MS}}$ value of $\sin^2\theta_W$, and
$\rho_{\nu e}$ and $\hat{\kappa}_{\nu e}$ denote process specific corrections. 
At zero-momentum transfer, these are $\rho_{\nu e}=1.0128$ and $\hat{\kappa}_{\nu e}=0.9963$ \cite{Erler:2012}.
Since the deviations of these parameters
from one are small compared to the sensitivity of Borexino, they,
and any flavor dependence that may exist for the $\alpha=e,\,\tau$ cases,
can be ignored.
In our analysis, we use the $\overline{\mathrm{MS}}$ value of
$\hat{s}^2_Z(0.862\,\mathrm{MeV})=0.2386$ \cite{Agarwalla:2010ty} 
for $\sin^2\theta_W$,
and the tree-level expressions for the couplings.
}

The presence of flavor-diagonal NSIs, $\varepsilon_{\alpha L/R}$,
will shift the coupling constants in the above expression to
%
%%%%%%%%%%%%%%%%%%%%%%%%%%%%%%%%%%%%%
\begin{equation}
g_{\alpha L} \;\rightarrow\; \tilde{g}_{\alpha L} = g_{\alpha L} + \varepsilon_{\alpha L}\;, \qquad
g_{\alpha R} \;\rightarrow\; \tilde{g}_{\alpha R} = g_{\alpha R} + \varepsilon_{\alpha R}\;.
\label{eqn2}
\end{equation}
%%%%%%%%%%%%%%%%%%%%%%%%%%%%%%%%%%%%%%
%
This interaction between electron and neutrino, with possible shifts in the coupling constants,
can be observed via the elastic scattering of a neutrino of flavor $\alpha$ 
off of an electron at rest, which has the differential cross section \cite{Vogel:1989iv}
%
%%%%%%%%%%%%%%%%%%%%%%%%%%%%%%%%%%%%%
\begin{equation}
\frac{d \tilde{\sigma}_{\nu_{\alpha}}(E_{\nu_\alpha},T)}{dT} = \frac{2G^2_{F} m_{e}}{\pi} 
\left[
\tilde{g}^2_{\alpha L} + \tilde{g}^2_{\alpha R}
\left(
1 - \dfrac{T}{E_{\nu_\alpha}}
\right)^2 - \tilde{g}_{\alpha L} \tilde{g}_{\alpha R} \dfrac{m_e T}{E_{\nu_\alpha}^2}
\right].
\label{eqn3}
\end{equation}
%%%%%%%%%%%%%%%%%%%%%%%%%%%%%%
%
Here, $m_e$ is the electron mass, $E_{\nu_\alpha}$ is the initial neutrino energy,
and $T$ is the kinetic energy of the recoil electron which has the range 
\begin{equation}
0 \;\le\; T \;\le\; T_{\max} \;=\; \dfrac{E_{\nu_\alpha}}{1+m_e/2E_{\nu_\alpha}}\;.
\label{Trange}
\end{equation}
If the incoming neutrino beam is mono-energetic, 
no convolution of Eq.~(\ref{eqn3}) with the neutrino energy spectrum is necessary.

The $\Be$ solar neutrinos are produced via the K-shell electron capture processes \cite{Tilley:2002vg}
\begin{equation}
\Be + e^- 
\;\rightarrow \;
\begin{cases}
\Li{7} + \nu_e\quad & (89.6\%) \;,\\
{\Li{7}}^*(0.48) + \nu_e\quad & (10.4\%) \;,
\end{cases}
\label{Be7Kcapture}
\end{equation}
yielding mono-energetic neutrinos of energy
$0.862\,\mathrm{MeV}$ and $0.384\,\mathrm{MeV}$, respectively. 
%$0.8618\,\mathrm{MeV}$ or $0.3842\,\mathrm{MeV}$. 
Borexino is sensitive to the $0.862\,\mathrm{MeV}$ component, for which the maximum
recoil energy is $T_{\max}=0.665\,\mathrm{MeV}$.
This component arrives at the Earth as a superposition of the three neutrino flavors due to 
the MSW effect and vacuum oscillation.
If we denote the survival probability of $\nu_e$ in this component
at the Borexino detector as $P_{ee}$,
then the total neutrino-electron scattering cross section there will be the combination
%
%%%%%%%%%%%%%%%%%%%%%%%%%%%%%%%%%%%%%
\begin{equation}
\frac{d \tilde{\sigma}_{\nu}(T)}{dT} 
\;=\; 
\left[
P_{ee}\,\frac{d \tilde{\sigma}_{\nu_e}(T)}{dT}
+(1-P_{ee})
\left(
c^2_{23}\,
\frac{d \tilde{\sigma}_{\nu_\mu}(T)}{dT}
+
s^2_{23}\,
\frac{d \tilde{\sigma}_{\nu_\tau}(T)}{dT}
\right)
\right],
\label{eqn4}
\end{equation}
%%%%%%%%%%%%%%%%%%%%%%%%%%%%%%
%
where $c^2_{23}=\cos^2\theta_{23}$ and $s^2_{23}=\sin^2\theta_{23}$.
Thus, the measurement of the $T$-dependence of the $\Be$ solar neutrino 
elastic scattering event rate
will let us constrain the values of $\tilde{g}_{\alpha L}$ and $\tilde{g}_{\alpha R}$
for all flavors.

An actual detector, however, cannot measure the recoil electron energy to arbitrary precision,
and one must take its finite energy resolution into account.
If we denote the probability of detecting energy $T_A$ for an electron with kinetic energy $T$
by $R(T_A,T)$, the differential cross section as a function of 
the actual detection energy $T_A$ is given by
\begin{equation}
\dfrac{d\overline{\sigma}_{\nu_\alpha}(T_A)}{dT_A}
\;=\; \int_{0}^{T_{\max}} R(T_A,T)\;\dfrac{d\tilde{\sigma}_{\nu_\alpha}(T)}{dT}\;dT\;.
\end{equation} 
For Borexino, we take the energy resolution function $R(T_A,T)$ to be a
gaussian with a $T$-dependent standard deviation
\begin{equation}
R(T_A,T) \;=\; \frac{1}{ \sqrt{2\pi}\,\sigma(T) }
\exp \left[-\frac{ (T_A-T)^2}{2\,[\,\sigma(T)\,]^2}\right],
\label{eqn:resolution}
\end{equation}
where $\sigma(T)$ is given by \cite{Bellini:1995wd}
\begin{equation}
\sigma(T) \;=\; \sigma_0 \left( \frac{T}{\rm MeV} \right)^{1/2}\;,\qquad
\sigma_0 \;=\; 50\,\mathrm{keV}\;.
\end{equation}
This will `blur-out' the shape of the energy spectrum somewhat, smoothing out the
Compton-like edge of $d\tilde{\sigma}_\nu(T)/dT$, but still retain its basic
overall shape.

Thus, for an incoming $0.862\,\mathrm{MeV}$ 
$\Be$ neutrino flux of $\Phi_{\Be}^{0.862}$ and number of electrons $N_e$ in the
fiducial volume of the detector, the number of recoil electrons 
detected with energy in the bin $T_1<T_A<T_2$ per unit time is given by
\begin{eqnarray}
\dfrac{dN(T_1,T_2)}{dt}
& = & N_e\Phi_{\Be}^{0.862}\int_{T_1}^{T_2} \dfrac{d\overline{\sigma}_{\nu}(T_A)}{dT_A}\;dT_A \cr
& = & 
N_e\Phi_{\Be}^{0.862}
\Biggl[
P_{ee}\int_{T_1}^{T_2} \dfrac{d\overline{\sigma}_{\nu_e}(T_A)}{dT_A}\;dT_A 
\cr
& &
\qquad\qquad +(1-P_{ee})(1-s^2_{23})
\int_{T_1}^{T_2} \dfrac{d\overline{\sigma}_{\nu_\mu}(T_A)}{dT_A}\;dT_A \cr
& & 
\qquad\qquad +(1-P_{ee})s^2_{23}
\int_{T_1}^{T_2} \dfrac{d\overline{\sigma}_{\nu_\tau}(T_A)}{dT_A}\;dT_A
\Biggr]
\;.
\label{NT1T2}
\end{eqnarray}
By measuring this spectrum, Borexino can constrain both $\varepsilon_{eL/R}$ and
$\varepsilon_{\tau L/R}$.
However, the precision of those constrains will depend on the uncertainty in the prefactor
$N_e\Phi_{\Be}^{0.862}$, which is still quite significant,
and those in $P_{ee}$ and $s_{23}^2$.

The measurement of the recoil electron energy spectrum in Borexino
is further complicated by the fact that
it is impossible to distinguish between electrons
from $\nu_\alpha e$ scattering and those from $\beta$-decay of radioactive
nuclei.  
The most significant $\beta$ backgrounds in Borexino are those from the decays
\begin{equation}
\begin{array}{rcll}
\Kr      & \;\longrightarrow\; & \Rb{85} + e^- + \overline{\nu}_e\quad 
& (Q=0.687\;\mathrm{MeV},\;t^{1/2}=10.756\;\mathrm{years},\;99.57\%)\;,\\
\Bi{210} & \longrightarrow & \Po{210} + e^- + \overline{\nu}_e\quad
& (Q=1.161\;\mathrm{MeV},\;t^{1/2}=5.012\;\mathrm{days},\;100\%)\;.
\end{array}
\label{Kr&BiDecays}
\end{equation}
%
%The energy spectrum of the electrons in these decays can be found in Ref.~\cite{XXX}.
As will be shown later, the electrons from $\Kr$ decay are particularly problematic.
Of the decay products, $\Rb{85}$ is stable while
$\Po{210}$ subsequently undergoes $\alpha$-decay
\begin{equation}
\Po{210} \; \longrightarrow \; \Pb{206} + \He{4}\qquad
(Q=5.3\;\mathrm{MeV},\;t^{1/2}=138\;\mathrm{days},\;100\%)\;.
\label{PoDecay}
\end{equation}
This $\alpha$-particle can mimic $\beta$-particles in the relevant energy range, but this
particular background can be removed reliably using pulse shape analysis \cite{Back:2007gp}.
%\footnote{Gatti's parameter \cite{Back:2007gp} take on positive values for $\alpha$-particles and negative values for $\beta$-particles.
%Its value is calculated by mean of an improvement of pulse-shape discrimination technique used to distinguish the output electronic signal form PMTs of $\beta$-events from that of $\alpha$-events. Plotting all events on this parameter, in paritcular way those in $\Po{210}$ peak energy range, they have two well-resolved Gaussian distributions with different mean values for respective kind of particles and, by means of fit of them, it's possible to reconstruct the beta spectrum by cutting away the $\Po{210}$ events.}
$\Pb{206}$ is stable.

In the next section, we will review the properties of the
Borexino detector, and how well these $\beta$-decay backgrounds are understood.

%%%%%%%%%%%%%%%%%%%%%%%%%%%%%%%%%%%%%%%%%%%%%%%%%%%%%%%%%%%%%%%%%%%%%%%%%%%%%%
%%%%%%%%%%%%%%%%%%%%%%%%%%%%%%%%%%%%%%%%%%%%%%%%%%%%%%%%%%%%%%%%%%%%%%%%%%%%%%
\section{The Borexino Detector and beta-decay Backgrounds}
\label{sec:borexino}
%%%%%%%%%%%%%%%%%%%%%%

Borexino is a real-time solar neutrino detector designed to measure the 
$0.862\,\mathrm{MeV}$ mono-energetic $\Be$ solar neutrinos.
It is situated in Hall~C of the Laboratori Nazionali del Gran
Sasso (LNGS), Italy, below 1400 meters of rock (3800 meters water equivalent)
where the muon flux is suppressed by a factor of $\sim\!10^{6}$ compared to the Earth's surface.
Borexino's spherical vessel is filled with $\sim$278 tons of liquid scintillator 
(pseudocumene doped with 2,5-diphenyloxazole as a wavelength shifter),
with the fiducial volume consisting of the central 100 tons.
In the period from May 16, 2007 to May 8, 2010, Borexino had
740.7 live days of data taking,
corresponding to $153.6\;\mathrm{ton\!\cdot\!years}$ of fiducial exposure.
\cite{Arpesella:2007xf,Arpesella:2008mt,Bellini:2011rx}

The neutrinos are detected via their elastic scattering off of electrons in the scintillator.
The scintillation light from the recoil electrons spread isotropically from the event location, and are 
detected by an array of $\sim$2200 photomultiplier tubes (PMTs) mounted onto a stainless steal sphere, 
looking into the detector volume.
A neutrino event is identified by multiple PMT hits within a Trigger Time Window (TTW) of $60$ nanoseconds,
with the total number of photoelectrons collected exceeding a given threshold. 
This triggers the recording of all PMT hits for $16$ microseconds, from which the electron
recoil energy $T_A$ is reconstructed.
The location of the event within the detector is determined from the difference in
arrival times of the TTW photons to different parts of the PMT array.
For a $T_A\sim 1\,\mathrm{MeV}$ event, this system can identify its location to a
precision of about 10~cm \cite{Alimonti:2008gc}.

The chemical composition of pseudocumene is $\mathrm{C}_9\mathrm{H}_{12}$ with atomic mass $120.19$,
and 66 electrons per molecule. 
Thus, the number of electrons within the fiducial volume of 100~tons of pseudocumene 
can be calculated to be $N_e = 3.307\times 10^{31}$. 
However, due to the limited event-position resolution of Borexino discussed above, 
there exists an uncertainty in the fiducial volume of ${}^{+0.5}_{-1.3}\,\%$ 
which propagates directly into an uncertainty in $N_e$ \cite{Bellini:2011rx}.

As discussed in the previous section, the electrons from $\Kr$ and $\Bi{210}$ $\beta$-decay
present significant backgrounds to the measurement of the $\Be$ signal. 
Due to the $Q$-values of these decays being close to the $\Be$ solar neutrino energy, 
their $\beta$-decay spectra overlap with the $\Be$ recoil electron spectrum
as shown in figure~\ref{fig:event}.
Of the other backgrounds present, 
that due to $\C{14}$, which $\beta$-decays with $Q$-value of $0.1565\,\mathrm{MeV}$,
is large but occupies a much lower energy range \cite{Alimonti:1998rc}.
The background due to cosmogenic $\C{11}$ \cite{Galbiati:2004wx}, which 
$\beta^+$-decays with $Q$-value of $0.96\,\mathrm{MeV}$, is at larger energies.
A fit to the measured count rate assuming SM interactions yields 
(TABLE~I of Ref.~\cite{Bellini:2011rx})
\begin{equation}
\begin{array}{rll}
\Be     
&\;\;:\;\; 46.0 \pm 1.5 (\mathrm{stat}) {}^{+1.5}_{-1.6} (\mathrm{syst})\quad 
& \mathrm{counts/(day \cdot 100\,tons)}\;, \\
\Kr      
&\;\;:\;\; 31.2 \pm 1.7 (\mathrm{stat}) \pm 4.7 (\mathrm{syst})\quad 
& \mathrm{counts/(day \cdot 100\,tons)}\;, \\
\Bi{210} 
&\;\;:\;\; 41.0 \pm 1.5 (\mathrm{stat}) \pm 2.3 (\mathrm{syst})\quad
& \mathrm{counts/(day \cdot 100\,tons)}\;,
\end{array}
\label{CountRateFit}
\end{equation}
showing that the count rates of the $\Kr$ and $\Bi{210}$ backgrounds are of the same order of magnitude 
as that of the $\Be$ signal.
Thus, the distortions of the shape of the count rate spectrum due to these backgrounds
are significant.
Furthermore, the shapes of the $\Kr$ and $\Bi{210}$ $\beta$-spectra are such that 
the said distortions are similar to those due to non-zero values of the NSI parameters.
Thus, in order to extract the bounds on the NSI parameters from the count rate data,
it is crucial that we can determine the $\Kr$ and $\Bi{210}$ backgrounds independently.

For $\Kr$, this is possible 
by utilizing the following decay chain which constitutes $0.43\%$ of 
$\Kr$-decay :
\begin{equation}
\begin{array}{rcll}
\Kr & \;\longrightarrow\; & \Rb{85\mathrm{m}} + e^- + \overline{\nu}_e\qquad 
& (Q=0.173\;\mathrm{MeV})\;, \\
\Rb{85\mathrm{m}} & \;\longrightarrow\; & \Rb{85} + \gamma
& (Q=0.514\;\mathrm{MeV},\,t^{1/2}=10^{-6}\,\mathrm{s})\;.
\end{array}
\label{Kr&RbDecay}
\end{equation}
Delayed coincidence measurements of the $\beta$ and $\gamma$ from this decay-chain
has yielded the total count rate of \cite{Ianni:2004tk}
\begin{equation}
\Kr\;\;:\;\; 30.4 \pm 5.3 (\mathrm{stat}) \pm 1.3 (\mathrm{syst})\;
\mathrm{counts/(day \cdot 100\,tons)}\;,
\label{KrDirect}
\end{equation}
which is consistent with the fit value listed in Eq.~(\ref{CountRateFit}).
We will use this direct measurement value as a constraint in our analysis.

$\Bi{210}$ is a pure $\beta$-emitter produced at the end of the $\Rn{222}$ decay chain:
\begin{equation}
\Rn{222}
\,\xrightarrow{\alpha}\,\Po{218}
\,\xrightarrow{\alpha}\,\Pb{214}
\,\xrightarrow{\beta}\,\Bi{214}
\,\xrightarrow{\beta}\,\Po{214}\\
\,\xrightarrow{\alpha}\,\Pb{210}
\,\xrightarrow{\beta}\,\Bi{210}
\,\xrightarrow{\beta}\,\Po{210}
\,\xrightarrow{\alpha}\,\Pb{206}\;.
\end{equation}
It was proposed in Ref.~\cite{Villante:2011zh} that the quantity of $\Bi{210}$ in liquid scintillation detectors
may be determined directly from the quantity of its decay product $\Po{210}$, 
which in turn can be determined by measuring its $\alpha$-decay rate.
Unfortunately, this method cannot be used due to temporal instabilities of the Borexino data.
However, in the narrow energy range above the $\Be$ shoulder and below the 
lower end of the $\C{11}$ $\beta^+$-decay spectrum,
aka the ``$\Be$ valley,'' the count rate is dominated by $\Bi{210}$ decay. 
The count rates from the other two components that contribute in this valley, 
namely the CNO and $pep$ solar neutrinos, are too low compared to that of $\Bi{210}$
to be measured with statistical significance.
Indeed, the contributions of CNO and $pep$ neutrinos are small
throughout the energy range of our analysis, with count rates
comparable to the statistical errors of the $\beta$-decay backgrounds \cite{Collaboration:2011nga}.
Therefore, though the precision is limited by statistics,
fitting to the count rate in the valley can constrain the $\Bi{210}$ background
independently of the fit to the NSI parameters.

%%%%%%%%%%%%%%%%%%%%%%%%%%%%%%%%%%%%%%%%%%%%%%%%%%%%%%%%%%%%%%%%%%%%%%%%%%%%%%
%%%%%%%%%%%%%%%%%%%%%%%%%%%%%%%%%%%%%%%%%%%%%%%%%%%%%%%%%%%%%%%%%%%%%%%%%%%%%%
\section{Uncertainty in the Standard Solar Model Neutrino Flux Prediction}
\label{sec:Solar_Model}
%%%%%%%%%%%%%%%%%%%%%%

The precisions of the neutrino fluxes predicted by the Standard Solar Model (SSM)
depend on the precisions of the physical input parameters.
In the past, the $\Be$ neutrino flux prediction suffered from a large uncertainty 
(12\% in Ref.~\cite{Bahcall:2004fg} from 2004)
stemming mostly from a large uncertainty in the cross section of the reaction
\begin{equation}
\He{3}+\He{4}\;\rightarrow\;\Be+\gamma
\end{equation}
at energies relevant to reactions in the Solar core.  
This cross section had been measured by shooting an $\alpha$-beam
into a gas $\He{3}$ target, and counting the number of produced $\Be$ by either
detecting the $\gamma$s in the above reaction (prompt method), or by measuring
the amount of accumulated $\Be$ later by counting the $\gamma$s from the decay
of $\Li{7}^*(0.48)$ produced in the second reaction of Eq.~(\ref{Be7Kcapture}) (activation method).
The large uncertainty was due to a disagreement between the results obtained via the
prompt and activation methods. 
Extrapolating the results down to the relevant energies lead to an additional uncertainty.
In the last decade, however, these uncertainties have been reduced considerably by the
LUNA experiment at the LNGS by a resolution of the discrepancy between the two methods, 
and the measurement of the cross section at lower energies than before 
\cite{Confortola:2007nq,Costantini:2008ub}.
As a result, the error in the $\Be$ neutrino flux due to this particular uncertainty 
has been reduced from 8\% to 2.8\% \cite{PenaGaray:2008qe}.

Unfortunately, solar metalicity has come into the picture as a new source of large uncertainty. 
The abundance of various elements inside the Sun is inferred from 
the measurements of their abundances in planets, chondrites\footnote{A kind of meteorite.}, and other bodies
in the Solar system, and from the analysis of absorption lines in the Solar photosphere spectrum.
The abundances inferred from the latter depend on the hydrodynamic model used for the photosphere.
A determination based on a 1D model of the solar atmosphere published in 1998 \cite{Grevesse:1998bj}
(GS98) has been used as input to the SSM for many years.
A new determination based on a state-of-the-art 3D model, and other improvements, was
published in 2009 \cite{Asplund:2009fu} (AGSS09).
The inferred values of the metalicity at the solar surface is
$(Z/X)_\odot=0.0229$ in GS98, but 
$(Z/X)_\odot=0.0178$ in AGSS09.\footnote{%
$X$, $Y$, and $Z$ respectively denote the mass fractions
of hydrogen, helium, and metals (elements other then H or He).}
The SSM predictions of the total $\Be$ neutrino flux 
(sum of the $0.862\,\mathrm{MeV}$ and $0.384\,\mathrm{MeV}$ components)
for the two cases are \cite{Serenelli:2011py}
\begin{eqnarray}
\Phi_{\Be}^{\mathrm{total}}
& = & 5.00\;(1\pm 0.07)\times 10^9\;\mathrm{cm^{-2}s^{-1}}\qquad (\mathrm{GS98})\;,\cr
& = & 4.56\;(1\pm 0.07)\times 10^9\;\mathrm{cm^{-2}s^{-1}}\qquad (\mathrm{AGSS09})\;.
\end{eqnarray}
The 7\% uncertainty in both cases are due to the uncertainties
in the cross sections of the reactions $\He{3}+\He{3}\rightarrow\He{4}+2p$ (2.5\%) and
$\He{3}+\He{4}\rightarrow\Be+\gamma$ (2.8\%), opacity (3.2\%), diffusion (2\%),
quoted uncertainties in the abundances (2\%), and various other sources \cite{CarlosTalk}.
As can be seen, the central value of the flux prediction 
can differ considerably depending on which set of abundances are adopted.
However, various predictions of the SSM based on the AGSS09 data set disagree with
solar properties inferred from helio-seismology, while those based on GS98 show
excellent agreement. (See e.g. Ref.~\cite{serenelli:2011ea} and references therein.)
%Since it is beyond the scope of this paper to resolve this disagreement,
Thus, we will simply use the GS98 based flux for our analysis, which was also the flux used
by Borexino to constrain $P_{ee}$ \cite{Bellini:2011rx}.
The $0.862\,\mathrm{MeV}$ neutrinos constitute $89.56\pm 0.04\%$ \cite{Tilley:2002vg} 
of the total flux, so the GS98 prediction of $\Phi_{\Be}^{0.862}$ will be
\begin{equation}
\Phi_{\Be}^{0.862}
\;=\; 4.48\;(1 \pm 0.07)\times 10^9\;\mathrm{cm^{-2}s^{-1}}\qquad (\mathrm{GS98})\;.
\label{GS98FluxPrediction}
\end{equation}
%

%%%%%%%%%%%%%%%%%%%%%%%%%%%%%%%%%%%%%%%%%%%%%%%%%%%%%%%%%%%%%%%%%%%%%%%%%%%%%%%
%\begin{table}[t]
%\begin{center}
%\begin{tabular}{|c|c|c|c|c||c|}
%\hline
%$\quad S_{33}\quad$ & $\quad S_{34}\quad$ & Opacity & Diffusion & Composition (Fe,C,O) & Total \\
%\hline
%$2.5\%$             & $2.8\%$             & $3.2\%$ & $2.0\%$   & $2.0\%$     & $7\%$ \\
%\hline
%\end{tabular}
%\caption{Error Budget of the $\Be7$ solar neutrino flux prediction of the SSM.
%The astrophysical $S$-factors $S_{33}$ and $S_{34}$ parameterize the cross sections
%of the reactions $\He{3}+\He{3}\rightarrow\He{4}+2p$ and
%$\He{3}+\He{4}\rightarrow\Be+\gamma$, respectively.
%}
%\label{Be7Errors}
%\end{center}
%\end{table}
%%%%%%%%%%%%%%%%%%%%%%%%%%%%%%%%%%%%%%%%%%%%%%%%%%%%%%%%%%%%%%%%%%%%%%%%%%%%%%

%%%%%%%%%%%%%%%%%%%%%%%%%%%%%%%%%%%%%%%%%%%%%%%%%%%%%%%%%%%%%%%%%%%%%%%%%%%%%%
%%%%%%%%%%%%%%%%%%%%%%%%%%%%%%%%%%%%%%%%%%%%%%%%%%%%%%%%%%%%%%%%%%%%%%%%%%%%%%
\section{Uncertainties in the Mixing Angles}
\label{sec:angles}

As discussed in section~\ref{sec:nue}, we need to understand the
uncertainties in $N_e\Phi_{\Be}^{0.862}$, $P_{ee}$, and $s_{23}^2$ that appear
in Eq.~(\ref{NT1T2}) to constrain $\varepsilon_{\alpha L/R}$ ($\alpha=e,\,\tau$).
In the previous sections, we have seen that $N_e$ has an uncertainty of
${}^{+0.5}_{-1.3}\,\%$, while $\Phi_{\Be}^{0.862}$ has an uncertainty of $\pm 7\%$,
so the uncertainty in the product $N_e\Phi_{\Be}^{0.862}$ is dominated by that
in $\Phi_{\Be}^{0.862}$ and we can assign to it an uncertainty of $\pm 7\%$.
Let us now look at the uncertainties in $P_{ee}$ and $s^2_{23}$.

%%%%%%%%%%%%%%%%%%%%%%%%%%%%%%%%%%%%%%%%%%%%%%%%%%%%%%%%%%%%%%%%%%%%%%%%%%%%%%
\begin{table}[b]
%\begin{center}
\scalebox{0.9}{
\begin{tabular}{|l||c|c|c|}
\hline
Reference
& Ref.~\cite{Tortola:2012te} 
& Ref.~\cite{Fogli:2012ua} 
& Ref.~\cite{GonzalezGarcia:2012sz}
\\
\hline\hline
$\Delta m_{21}^2 (10^{-5}\mathrm{eV}^2)$
& $7.62\pm 0.19$ 
& $7.54^{+0.26}_{-0.22}$
& $7.50^{+0.205}_{-0.160}$
\\
\hline
$\Delta m_{31}^2 (10^{-3}\mathrm{eV}^2)$ (N)
& $2.55^{+0.06}_{-0.09}$ 
& $2.43^{+0.06}_{-0.10}$
& $2.49^{+0.055}_{-0.051}$
\\
$\Delta m_{13}^2 (10^{-3}\mathrm{eV}^2)$ (I)
& $2.43^{+0.07}_{-0.06}$ 
& $2.42^{+0.07}_{-0.11}$
& $\Delta m_{23}^2 (10^{-3}\mathrm{eV}^2) = 2.47^{+0.064}_{-0.073}$
\\
\hline
$\sin^2\theta_{13}$ (N)
& $0.0246^{+0.0029}_{-0.0028}$
& $0.0241\pm 0.0025$
& $0.025 \pm 0.0023$
\\
$\sin^2\theta_{13}$ (I)
& $0.0250^{+0.0026}_{-0.0027}$
& $0.0244^{+0.0023}_{-0.0025}$
& 
\\
\hline
$\sin^2\theta_{23}$ (N)
& $0.427^{+0.034}_{-0.027}\oplus 0.613^{+0.022}_{-0.040}$
& $0.386^{+0.024}_{-0.021}$
& $0.41^{+0.030}_{-0.029}\oplus 0.60^{+0.020}_{-0.026}$
\\
$3\sigma$ range
& $0.36\rightarrow 0.68$
& $0.331\rightarrow 0.637$
& $0.34\rightarrow 0.67$
\\
\cline{1-3}
$\sin^2\theta_{23}$ (I)
& $0.600^{+0.026}_{-0.031}$
& $0.392^{+0.039}_{-0.022}$
&
\\
$3\sigma$ range
& $0.37\rightarrow 0.67$ 
& $0.335\rightarrow 0.663$
&
\\
\hline
$\sin^2\theta_{12}$
& $0.320{}^{+0.016}_{-0.017}$
& $0.307{}^{+0.018}_{-0.016}$
& $0.31\pm 0.013$
\\
\hline\hline
$P_{ee}$
& $0.565^{+0.013}_{-0.011}$
& $0.574\pm 0.013$ 
& $0.572\pm 0.010$
\\
\hline
\end{tabular}
}
\caption{$1\sigma$ bounds on the neutrino mixing angles and mass-squared differences
from the global fits performed in
Refs.~\cite{Tortola:2012te}, \cite{Fogli:2012ua}, and \cite{GonzalezGarcia:2012sz},
and the corresponding $\nu_e$ survival probability $P_{ee}$ for the $\Be$ neutrinos.
N and I stand for normal and inverted hierarchies.
The numbers cited from Ref.~\cite{GonzalezGarcia:2012sz} are those obtained
by assuming reactor neutrino fluxes of Huber \cite{Huber:2011wv},
with the mass hierarchy marginalized for the mixing angle values. 
The CP violating phase $\delta_{\mathrm{CP}}$ is essentially unconstrained 
so the fit values are not shown.
The agreement is good for all parameters except $\sin^2\theta_{23}$,
the $\chi^2$ of which has two local minima near $0.4$ and $0.6$.
These minima are roughly degenerate, separated by only a minuscule `bump' between them,
and which one is preferred depends on the mass hierarchy and minute details of the 
global analyses.
}
\label{sin2thetaXY} 
%\end{center}
\end{table}
%%%%%%%%%%%%%%%%%%%%%%%%%%%%%%%%%%%%%%%%%%%%%%%%%%%%%%%%%%%%%%%%%%%%%%%%%%%%%%

The $\nu_e$ survival probability $P_{ee}$ is dependent on the neutrino energy $E_\nu$.
At $E_\nu=0.862\,\mathrm{MeV}$, matter effects are negligible and we can approximate
(see, for instance, Ref.~\cite{Nakamura:2012})
\begin{equation}
P_{ee} \;\approx\; 1 - \frac{1}{2}\sin^2(2\theta_{12})\;.
\end{equation}
The best-fit values of the neutrino mixing angles from three recent global fits,
Refs.~\cite{Tortola:2012te}, \cite{Fogli:2012ua}, and \cite{GonzalezGarcia:2012sz},
are listed in Table~\ref{sin2thetaXY}.
These analyses are more recent that what was used in the 
2012 Review of Particle Properties \cite{Beringer:2012}, and include new results
from Super-Kamiokande \cite{Itow:2012}, MINOS \cite{Nichol:2012}, and other experiments
that were announced at the Neutrino 2012 conference in Kyoto (June 3--9, 2012).

The agreement in the values of $\sin^2\theta_{12}$ from the three fits is good, 
and rounding up the resulting $P_{ee}$ to the second decimal place yields
\begin{equation}
P_{ee} \;=\; 0.57\pm 0.01
\end{equation}
for all three. We will use this value for $P_{ee}$ in our analysis.\footnote{%
The value based on the Borexino data assuming Eq.~(\ref{GS98FluxPrediction}) is $P_{ee}=0.51\pm 0.07$ \cite{Bellini:2011rx}.}
Since the factors $P_{ee}$ and $(1-P_{ee})$ in Eq.~(\ref{NT1T2}) appear
multiplied by the factor $N_e\Phi_{\Be}^{0.862}$ with a $\pm 7\%$ uncertainty,
the small uncertainties in $P_{ee}$ and $(1-P_{ee})$ can be ignored.

The value of $\sin^2\theta_{23}$ is problematic.
While the new Super-Kamiokande atmospheric neutrino data still favors the maximal
$\sin^2 2\theta_{23}=1.00\,(\ge 0.94\,(90\%\,\mathrm{C.L.}))$ 
as the best fit value in a two-flavor oscillation
analysis \cite{Itow:2012}, preliminary results from MINOS prefer 
$\sin^2 2\theta_{23}=0.94^{+0.04}_{-0.05}$, leading to a combined
value of $\sin^2 2\theta_{23}=0.96\pm 0.04$ \cite{Nichol:2012}, 
which corresponds to either $\sin^2\theta_{23}=0.4$ or $0.6$.
This degeneracy is broken in a full three flavor oscillation analysis of the
atmospheric neutrino data,
with the preferred value depending on the mass hierarchy \cite{GonzalezGarcia:2012sz,Itow:2012}.
However, as can be seen from Table.~\ref{sin2thetaXY}, 
the fits of Refs.~\cite{Tortola:2012te} and \cite{Fogli:2012ua}
do not agree on which value should be preferred over the other
even when the mass hierarchy is fixed. 
Ref.~\cite{GonzalezGarcia:2012sz} marginalizes over the mass hierarchy, so the
degeneracy remains.
One also needs to be aware that
a small shift in the preferred value of $\sin^2 2\theta_{23}$ could
shift the preferred values of $\sin^2\theta_{23}$ considerably.
So a conservative assessment of the current situation would be that
the value of $\sin^2\theta_{23}$ is somewhere in the $3\sigma$ range
of $0.34\rightarrow 0.67$ (taken from Ref.~\cite{GonzalezGarcia:2012sz}),
without a particularly preferred value.
Thus, in this analysis, we choose to use 
the central $\sin^2\theta_{23}=0.5$ as the reference value,
and associate with it an uncertainty obtained by dividing the width of the $3\sigma$
range by six:
\begin{equation}
s_{23}^2 \;=\; \sin^2\theta_{23} \;=\; 0.500\pm 0.055\;.
\end{equation}
That is, we associate an 11\% uncertainty to $s_{23}^2$.

%%%%%%%%%%%%%%%%%%%%%%%%%%%%%%%%%%%%%%%%%%%%%%%%%%%%%%%%%%%%%%%%%%%%%%%%%%%%%%
%%%%%%%%%%%%%%%%%%%%%%%%%%%%%%%%%%%%%%%%%%%%%%%%%%%%%%%%%%%%%%%%%%%%%%%%%%%%%%
%\newpage
\section{Dependence of the Event Spectrum on the NSI Parameters}
\label{sec:event-nsi}
%%%%%%%%%%%%%%%%%%%%%%

%%%%%%%%%%%%%%%%%%%%%%%%%%%%%%%%%%%%%%%%%%%%%%%%%%%%%%%%%%%%%%%%%%%%%%%%%%%%%%
\begin{figure}[t]
\includegraphics[width=0.46\textwidth]{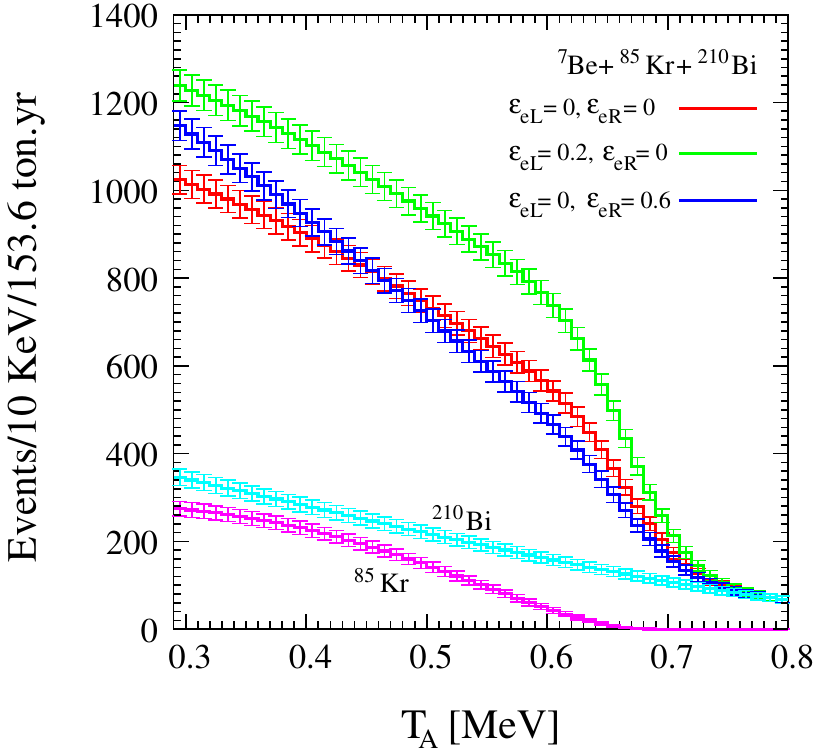}\hspace{1cm}
\includegraphics[width=0.46\textwidth]{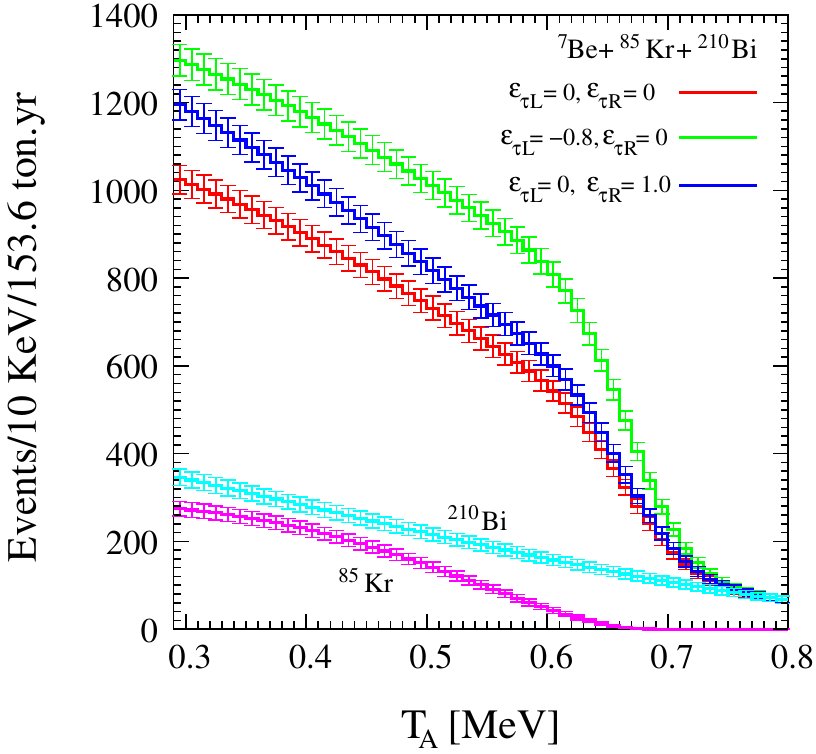}
\caption{\label{fig:event} 
Expected Borexino event spectrum for a fiducial exposure of $153.6\,\mathrm{ton{\cdot}year}$. 
The numbers shown are for $10\,\mathrm{keV}$ wide bins, with associated statistical errors.
The left-hand panel depicts the total number of events ($\Be+\Kr+\Bi{210}$) 
for different choices of $\varepsilon_{eL}$ and $\varepsilon_{eR}$ as shown in the legends.
The contributions from $\Kr$ and $\Bi{210}$, which are unaffected by the NSI parameters,
are also shown separately. 
The right-hand panel portrays the same for different choices of 
$\varepsilon_{\tau L}$ and $\varepsilon_{\tau R}$.
Note that we have taken the sample non-zero values of the NSI parameters to be
fairly large to exaggerate their effects.}
\end{figure}
%%%%%%%%%%%%%%%%%%%%%%%%%%%%%%%%%%%%%%%%%%%%%%%%%%%%%%%%%%%%%%%%%%%%%%%%%%%%%%

Before we proceed to our analysis of the Borexino $153.6\;\mathrm{ton{\cdot}year}$ data,
let us take a look at how the expected event spectrum depends on the NSI parameters
to obtain a feel on which features of the spectrum are relevant in constraining which parameter.

In Fig.~\ref{fig:event}, we show the expected $153.6\,\mathrm{ton{\cdot}year}$
Borexino event spectrum in the energy range $0.29\,\mathrm{MeV}<T_A<0.80\,\mathrm{MeV}$, 
in $10\,\mathrm{keV}$ wide bins, for several choices of the NSI parameters.
This energy range is dominated by the $\Be$, $\Kr$, and $\Bi{210}$ events, and all other
solar neutrinos and background contributions can be neglected.
The $\Be$ signal has been calculated with inputs
$\sin^2\theta_W=0.2386$, $P_{ee}=0.57$, $s^2_{23}=c^2_{23}=0.5$, and
$\Phi_{\Be}^{0.862}=4.48\times 10^9\,\mathrm{cm^{-2}s^{-1}}$,
as discussed in previous sections.
The total number of $\Be$ counts in the shown range is 14350 for the SM case
(all NSI parameters set to zero).
The $\Kr$ background has been fixed to the rate in Eq.~(\ref{KrDirect}), and
the $\Bi{210}$ background to the rate in Eq.~(\ref{CountRateFit}).
The total number of counts in the range is
5813 (10057) for the $\Kr$ ($\Bi{210}$) background.

It is quite evident from Fig.~\ref{fig:event} that 
the left-handed couplings, $\varepsilon_{\alpha L} (\alpha = e, \tau)$ affect the overall normalization, 
whereas the right-handed couplings, $\varepsilon_{\alpha R} (\alpha = e, \tau)$,
cause changes in both shape and normalization.  
Thus any uncertainty in the normalization of the $\Be$ signal
can severely deteriorate the sensitivity to $\varepsilon_{\alpha L}$, 
and also mimic the presence of $\varepsilon_{\alpha R}$ to some extent. 
The $\Kr$ background starts around the Compton-like edge of the $\Be$ signal, and
its presence affects the slope of the total event spectrum below this edge
in a way similar to non-zero values of $\varepsilon_{\alpha R}$.
Thus, the uncertainty in this background can be
expected to deteriorate the sensitivity to $\varepsilon_{\alpha R}$.
Note also that the events above the $\Be$ Compton-like edge,
$0.7\,\mathrm{MeV}\alt T_A$, are dominated by
the $\Bi{210}$ background as discussed previously,
so this background will be constrained 
by the events in this region when performing a fit.

%%%%%%%%%%%%%%%%%%%%%%%%%%%%%%%%%%%%%%%%%%%%%%%%%%%%%%%%%%%%%%%%%%%%%%%%%%%%%%
%\section{Correlations among NSIs and Uncertainty in $\Be$ and $\Kr$ rates}
%\label{sec:correlation}
%%%%%%%%%%%%%%%%%%%%%%%%%%%%%%%%%%%%%%%%%%%%%%%%%%%%%%%%%%%%%%%%%%%%%%%%%%%%%%

%%%%%%%%%%%%%%%%%%%%%%%%%%%%%%%%%%%%%%%%%%%%%%%%%%%%%%%%%%%%%%%%%%%%%%%%%%%%%%
\begin{figure}[tp]
\includegraphics[width=0.5\textwidth]{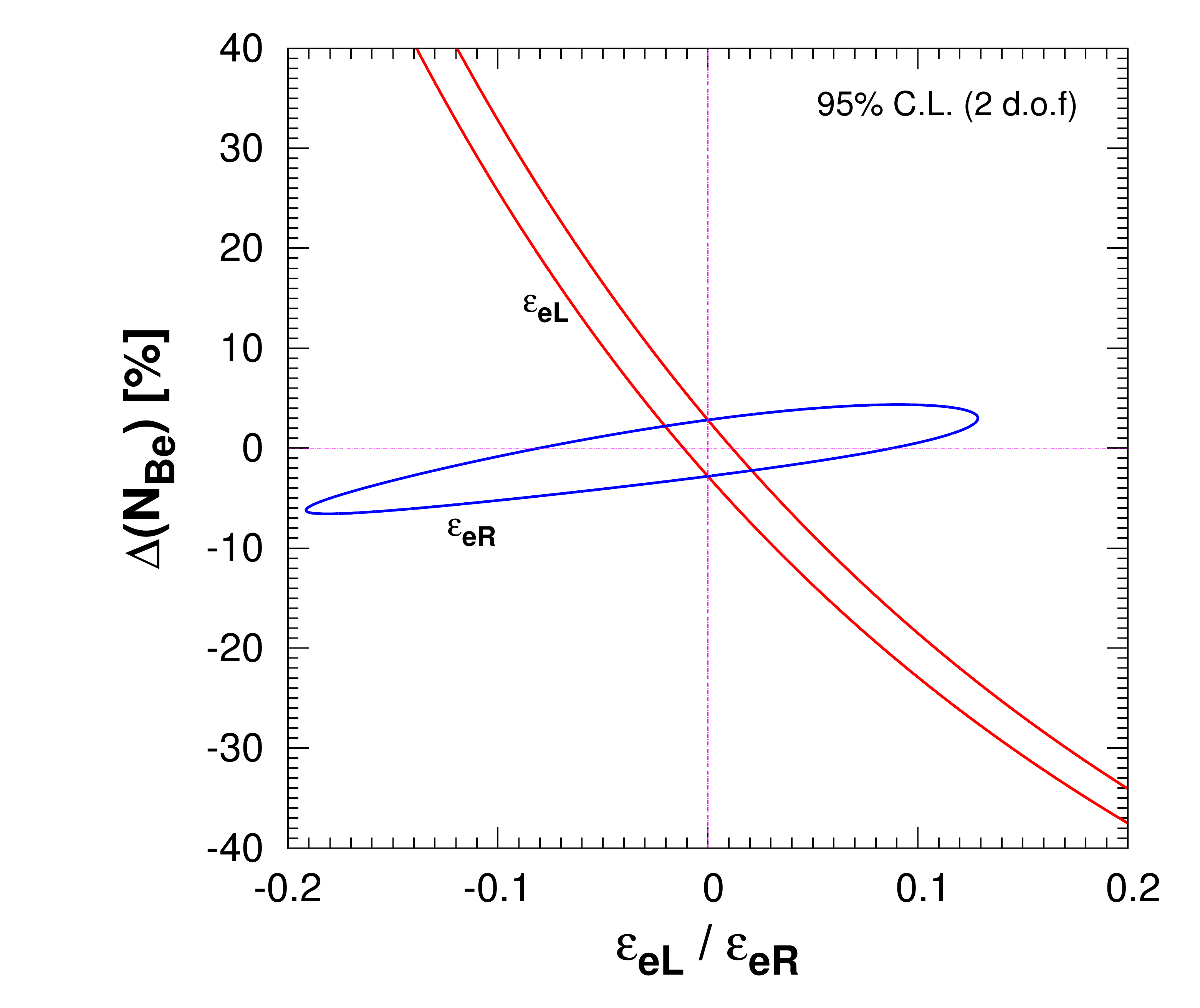}
\includegraphics[width=0.5\textwidth]{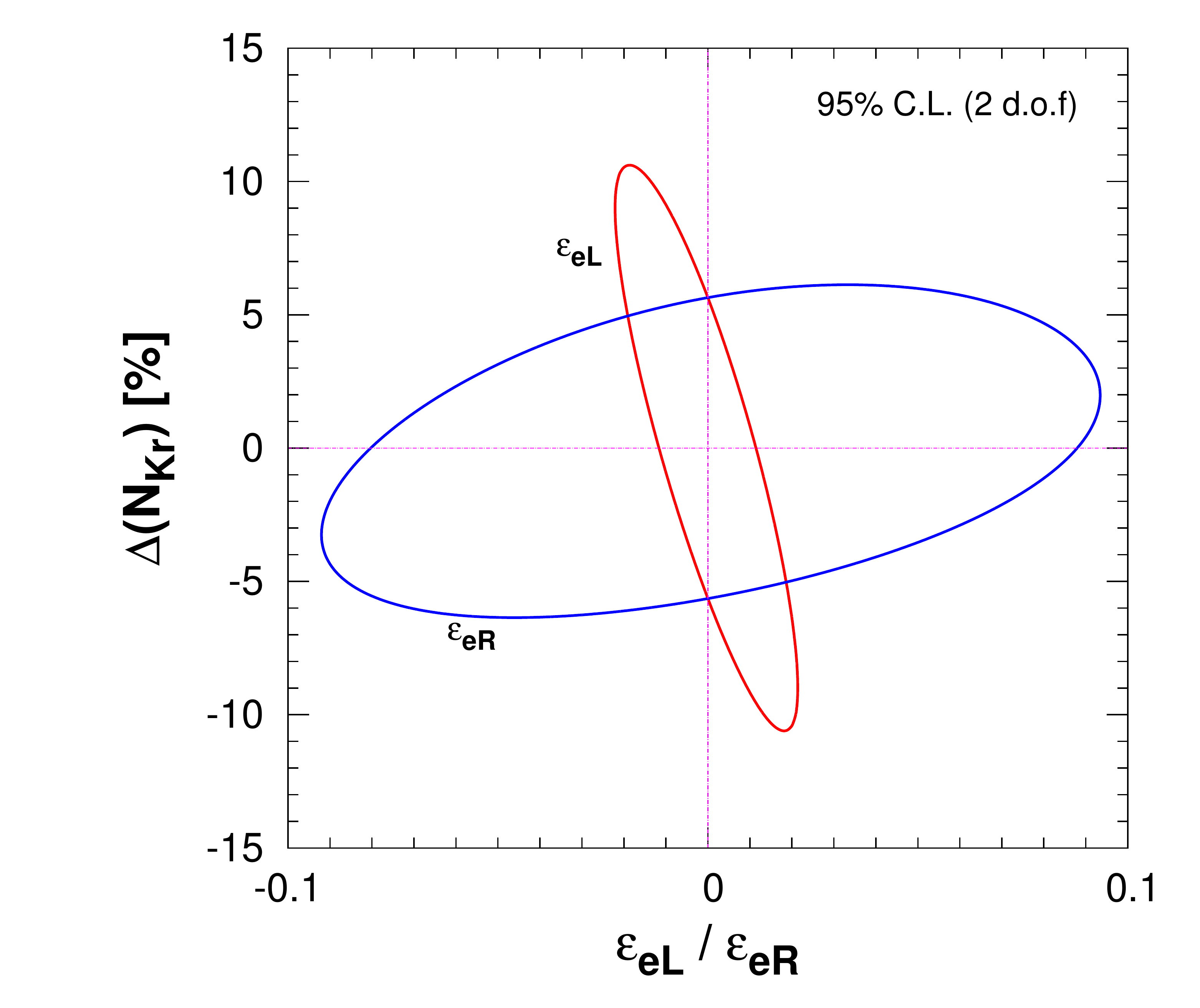}
\caption{\label{fig:param-correlation} 
Correlations between the NSI parameters and the normalizations of the $\Be$ signal and $\Kr$ background.
The left-hand panel shows the
95\% C.L. contours (2 d.o.f., $\Delta\chi^2=5.99$)  
in the $\Delta N_{\mathrm{Be}}$-$\varepsilon_{eL,R}$ plane,
while the right-hand panel shows the same
in the $\Delta N_{\mathrm{Kr}}$-$\varepsilon_{e L,R}$ plane.
The area inside the contours are allowed.
All implicit parameters are set to their reference values used in Fig.~\ref{fig:event}.
} 
\end{figure}
%%%%%%%%%%%%%%%%%%%%%%%%%%%%%%%%%%%%%%%%%%%%%%%%%%%%%%%%%%%%%%%%%%%%%%%%%%%%%%

To see how the uncertainties in the normalization of the $\Be$ signal
and the $\Kr$ background could affect the bounds on the NSI parameters,
we perform the following analysis.
Let $N_i^{\mathrm{th}}$ be the expected number of events in the $i$-th energy bin.
Construct the $\chi^2$ by
\begin{equation}
\chi^2(\varepsilon_{\alpha C},\Delta N_{\mathrm{X}}) 
\;=\; \sum_{i}
\dfrac{\bigl[\,
 N_i^{\mathrm{th}}(0,0)
-N_i^{\mathrm{th}}(\varepsilon_{\alpha C},\Delta N_{\mathrm{X}})
\,\bigr]^2}
{N_i^{\mathrm{th}}(0,0)}
\;,\qquad
\alpha\,=\,\mbox{$e$ or $\tau$}\;,
\end{equation}
where $C=L$ or $R$, and $\mathrm{X}=\mathrm{Be}$ or $\mathrm{Kr}$.
$\Delta N_\mathrm{Be}$ and $\Delta N_\mathrm{Kr}$ are respectively the percentage changes
in the normalizations of the $\Be$ signal and $\Kr$ background from their reference values.
Implicit parameters are all fixed to the reference values we used above.
This $\chi^2$ quantifies the ability of Borexino to distinguish between
the $(\varepsilon_{\alpha C},\Delta N_{\mathrm{X}})=(0,0)$ and non-zero cases.
Using this $\chi^2$ for the case $\alpha=e$, we plot the 95\% C.L. contours ($\Delta\chi^2=5.99$) in the
$\Delta N_{\mathrm{X}}$-$\varepsilon_{e C}$ planes in Fig.~\ref{fig:param-correlation}.

The left-hand panel of Fig.~\ref{fig:param-correlation} shows the correlations between
$\varepsilon_{e L/R}$ and $\Delta N_{\mathrm{Be}}$.
For $\varepsilon_{eL}$, we can see that it has a strong negative correlation to 
$\Delta N_{\mathrm{Be}}$, as was expected from our discussion above.
Thus, a large uncertainty in $\Delta N_{\mathrm{Be}}$ would 
lead to a large uncertainty in $\varepsilon_{eL}$.
$\varepsilon_{eR}$, on the other hand, is only weakly correlated with
$\Delta N_{\mathrm{Be}}$, again as expected.

The right-hand panel of Fig.~\ref{fig:param-correlation} shows the correlations between
$\varepsilon_{e L/R}$ and $\Delta N_{\mathrm{Kr}}$.
For $\varepsilon_{eL}$, we see that it is only weekly correlated with
$\Delta N_{\mathrm{Kr}}$ as expected.
$\varepsilon_{eR}$, however, is also only weakly correlated with
$\Delta N_{\mathrm{Kr}}$ somewhat contrary to expection.
Consequently, a reduction in the uncertainty of $\Delta N_{\mathrm{Kr}}$
will not lead to any significant reduction in the uncertainty of $\varepsilon_{eR}$.
In the following, we will find that the reduction of the $\Kr$ background itself,
and not just its uncertainty, is necessary to improve the bound on $\varepsilon_{eR}$.

%%%%%%%%%%%%%%%%%%%%%%%%%%%%%%%%%%%%%%%%%%%%%%%%%%%%%%%%%%%%%%%%%%%%%%%%%%%%%%
%%%%%%%%%%%%%%%%%%%%%%%%%%%%%%%%%%%%%%%%%%%%%%%%%%%%%%%%%%%%%%%%%%%%%%%%%%%%%%
\section{Analysis Results}
\label{sec:results}
%%%%%%%%%%%%%%%%%%%%%%

%%%%%%%%%%%%%%%%%%%%%%%%%%%%%%%%%%%%%%%%%%%%%%%%%%%%%%%%%%%%%%%%%%%%%%%%%%%%%%
\subsection{Method of Analysis}
\label{sec:setup}

Let us now proceed to our analysis of the Borexino $153.6\;\mathrm{ton{\cdot}year}$ data.
As in the previous section, 
we consider events in the energy range $0.29\,\mathrm{MeV} < T_A < 0.80\,\mathrm{MeV}$,
which we divide into $10\,\mathrm{keV}$ bins.
%This range is dominated by the $\Be$ signal, and the $\Kr$ and $\Bi{210}$ background events.
%All other solar neutrino contributions and backgrounds are negligible.

Let the number of measured counts in the $i$-th bin be $N_i^{\mathrm{exp}}$,
and its theoretical value $N_{i}^\mathrm{th}(\vec{\lambda})$,
where $\vec{\lambda}$ denotes the fit parameters that will be varied:
\begin{equation}
\vec{\lambda} \;=\; \bigl\{ 
\varepsilon_{e L},\,\varepsilon_{e R},\,
\varepsilon_{\tau L},\,\varepsilon_{\tau R},\,
\Delta N_{\mathrm{Be}},\Delta N_{\mathrm{Kr}},\Delta N_{\mathrm{Bi}},s_{23}^2
\bigr\}
\;.
\end{equation}
$\Delta N_{\mathrm{Be}}$, $\Delta N_{\mathrm{Kr}}$ and $\Delta N_{\mathrm{Bi}}$ respectively denote 
the percentage change in the $\Be$, $\Kr$, and $\Bi{210}$ event normalizations from their reference values.
The first is varied to account for the $\pm 7\%$ uncertainty 
in the $\Be$ neutrino flux,
while the second for the $\pm 18\%$ uncertainty in the $\Kr$ background, c.f. Eq.~(\ref{KrDirect}).
$s_{23}^2$ is varied with a $\pm 11\%$ uncertainty around the reference
value of $0.5$ as discussed in section~\ref{sec:angles}.
The $\chi^2$ is then defined as
\begin{equation}
\chi^2(\vec{\lambda}) \;=\; \sum_{i}\dfrac{\bigl[\,N_i^{\mathrm{exp}}-N_i^{\mathrm{th}}(\vec{\lambda})\,\bigr]^2}{N_i^{\mathrm{exp}}}
\;+\; \left[\dfrac{\Delta N_{\mathrm{Be}}}{7\%}\right]^2
\;+\; \left[\dfrac{\Delta N_{\mathrm{Kr}}}{18\%}\right]^2
+\left(\dfrac{s_{23}^2-0.5}{0.055}\right)^2
\;.
\end{equation}
No prior constraint is imposed on $\Delta N_{\mathrm{Bi}}$, which will be left for the fit to determine.

Since we do not have access to the raw $153.6\;\mathrm{ton{\cdot}year}$ Borexino data,
we reconstruct the counts $N_i^{\mathrm{exp}}$ from the fits listed in Eq.~(\ref{CountRateFit})
as a sum of $\Be$, $\Kr$, and $\Bi{210}$ events.
Contributions from other solar neutrinos and backgrounds are neglected.
The theoretical value $N_i^{\mathrm{th}}(\vec{\lambda})$ is calculated
with the parameter selection used in the previous section.
Note that $N_i^{\mathrm{exp}}$ is not equal to $N_i^{\mathrm{th}}(\vec{0})$.
Thus, the minimal value of $\chi^2$ will be non-zero:
\begin{equation}
\chi^2(\vec{\lambda})\;=\;\chi^2_{\min} + \Delta\chi^2(\vec{\lambda})\;.
\end{equation}
Using this $\Delta\chi^2$, we first place constrains on the four NSI parameters 
one at a time, keeping the other three NSI parameters zero, while
marginalizing\footnote{%
For a discussion on the `marginalization' procedure see, for instance, Refs.~\cite{Huber:2002mx} and \cite{Fogli:2002au}.}
over the three normalization parameters and $s_{23}^2$.
Then, we place constraints on pairs of NSI parameters, one flavor at a time,
keeping the other flavor pair zero, while again marginalizing over 
the normalization parameters and $s_{23}^2$.

%%%%%%%%%%%%%%%%%%%%%%%%%%%%%%%%%%%%%%%%%%%%%%%%%%%%%%%%%%%%%%%%%%%%%%%%%%%%%%
\subsection{One NSI parameter at a time limits}
\label{sec:one-parameter}

%%%%%%%%%%%%%%%%%%%%%%%%%%%%%%%%%%%%%%%%%%%%%%%%%%%%%%%%%%%%%%%%%%%%%%%%%%%%%%
\begin{figure}[t]
\includegraphics[width=0.45\textwidth]{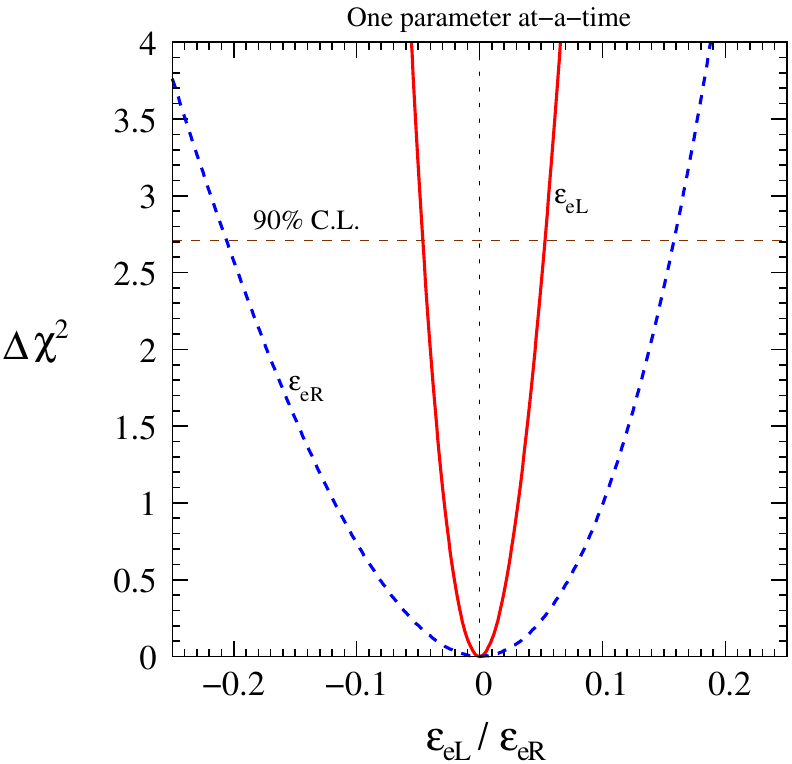}\hspace{1cm}
\includegraphics[width=0.45\textwidth]{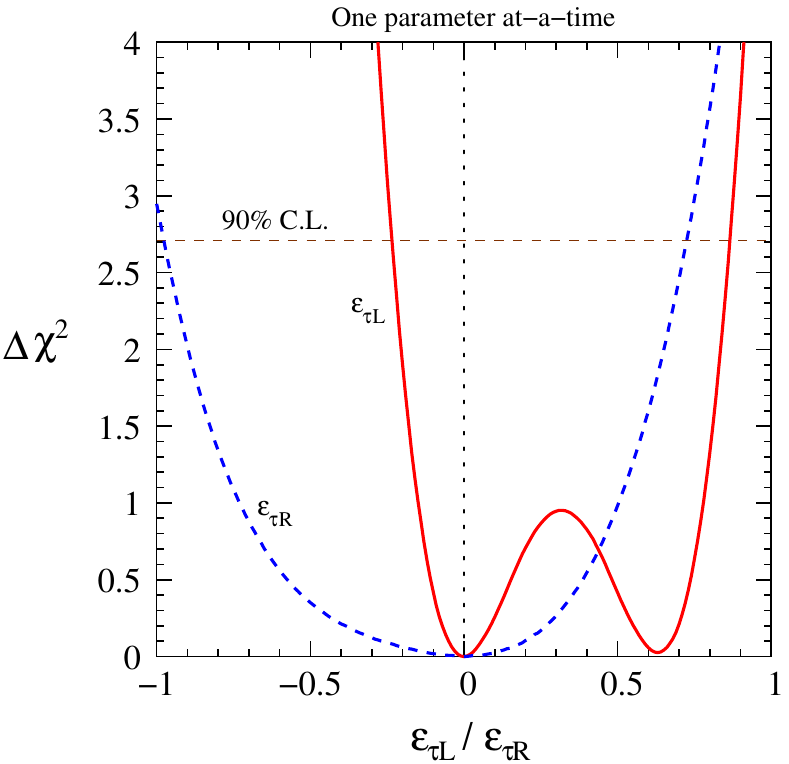}
\caption{\label{fig:one-parameter-at-a-time} 
Left-hand panel depicts the one NSI parameter at a time limit for $\varepsilon_{eL}$ and $\varepsilon_{eR}$. 
Right-hand panel shows the same for $\varepsilon_{\tau L}$ and $\varepsilon_{\tau R}$. 
The corresponding limits at 90\% C.L. (1 d.o.f, $\Delta\chi^2 = 2.71$) are 
listed in Table~\ref{tab:limits}.
The central values of the NSI parameters are zero despite our
using the value $P_{ee}=0.57$ in our reference model 
instead of $P_{ee}=0.51$ preferred by the Borexino data,
since the discrepancy is absorbed into the uncertainty in the $\Be$ flux. 
}
\end{figure}
%%%%%%%%%%%%%%%%%%%%%%%%%%%%%%%%%%%%%%%%%%%%%%%%%%%%%%%%%%%%%%%%%%%%%%%%%%%%%%

%%%%%%%%%%%%%%%%%%%%%%%%%%%%%%%%%%%%%%%%%%%%%%%%%%%%%%%%%%%%%%%%%%%%%%%%%%%%%%
\begin{table}[t]
\begin{center}
\scalebox{0.95}{
\begin{tabular}{|c||c|c|c|c|} 
\hline
& $\varepsilon_{eL}$ &$\varepsilon_{eR}$ &$\varepsilon_{\tau L}$ &$\varepsilon_{\tau R}$ \\
\hline\hline
$\phantom{\Big|}$ \ This work\ \ 
& $[-0.046,\;0.053\,]$ & $[-0.206,\;0.157\,]$ & $[-0.231,\;0.866\,]$ & $[-0.976,\;0.726\,]$ \\
%($153.6\;\mathrm{ton\!\cdot\!year}$) & & & & \\
\hline
$\phantom{\Big|}$Global limits \cite{Barranco:2007ej}\ 
& $[-0.03,\;0.08\,]$ & $[\,0.004,\;0.151\,]$ & $[-0.5,\;0.2\,]$ & $[-0.3,\;0.4\,]$ \\
\hline
\end{tabular}
}
\caption{\label{tab:limits} 
The 90\% C.L. limits on the flavor-diagonal
NSI parameters $\varepsilon_{eL}$, $\varepsilon_{eR}$, $\varepsilon_{\tau L}$
and $\varepsilon_{\tau R}$ based on $153.6\;\mathrm{ton\!\cdot\!years}$
of Borexino data.
In each case, only one NSI parameter and the three normalization parameters are allowed to float,
while the remaining three NSI parameters are fixed to zero.
The three normalization parameters are marginalized to obtain these bounds.
The second row lists the global bounds from Ref.~\cite{Barranco:2007ej} for comparison.}
\end{center}
\end{table}
%%%%%%%%%%%%%%%%%%%%%%%%%%%%%%%%%%%%%%%%%%%%%%%%%%%%%%%%%%%%%%%%%%%%%%%%%%%%%%

The dependence of $\Delta\chi^2$ on one NSI parameter, with the other three fixed to zero,
and after marginalization of the three normalization parameters, is show in  
Fig.~\ref{fig:one-parameter-at-a-time} for all four choices of the NSI parameter.
The corresponding 90\% C.L. limits (1 d.o.f, $\Delta\chi^2=2.71$) are listed in
Table~\ref{tab:limits}, together with the global fit values from Ref.~\cite{Barranco:2007ej}
for comparison.
We can see that the bounds based on Borexino data alone is already competitive with the
global fit to reactor+accelerator data.

Since July 2010, the Borexino experiment have undertaken a series of purification campaigns to reduce the radioactive backgrounds. 
The method of Nitrogen stripping has been quite successful in reducing the $\Kr$ background to 
roughly 30\% of previous levels \cite{Pallavicini:2012}. 
In 2012, with these lower backgrounds, the Borexino experiment has entered into its Phase~II run 
and continues to accumulate more data.

\begin{figure}[t]
\includegraphics[width=0.325\textwidth]{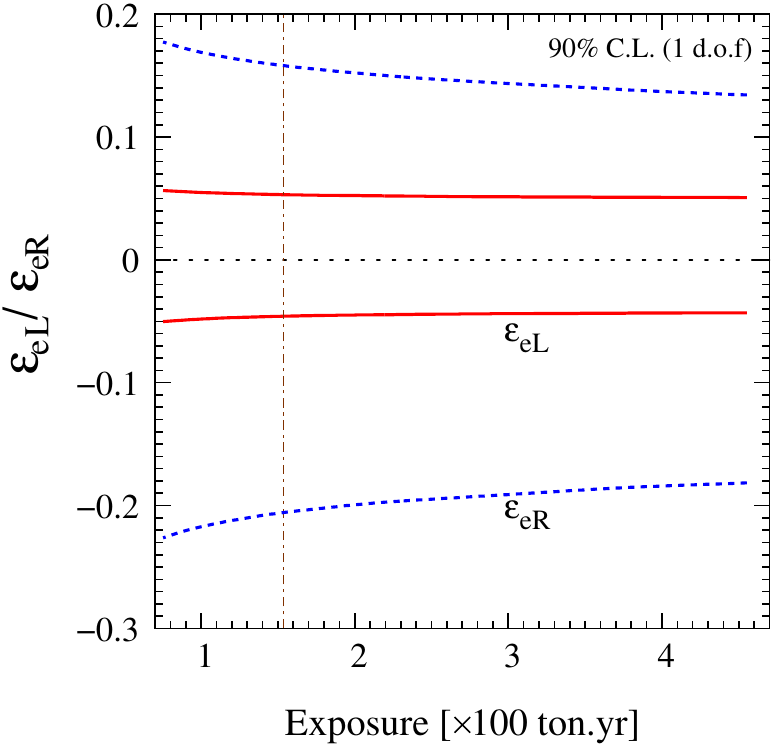}
\includegraphics[width=0.325\textwidth]{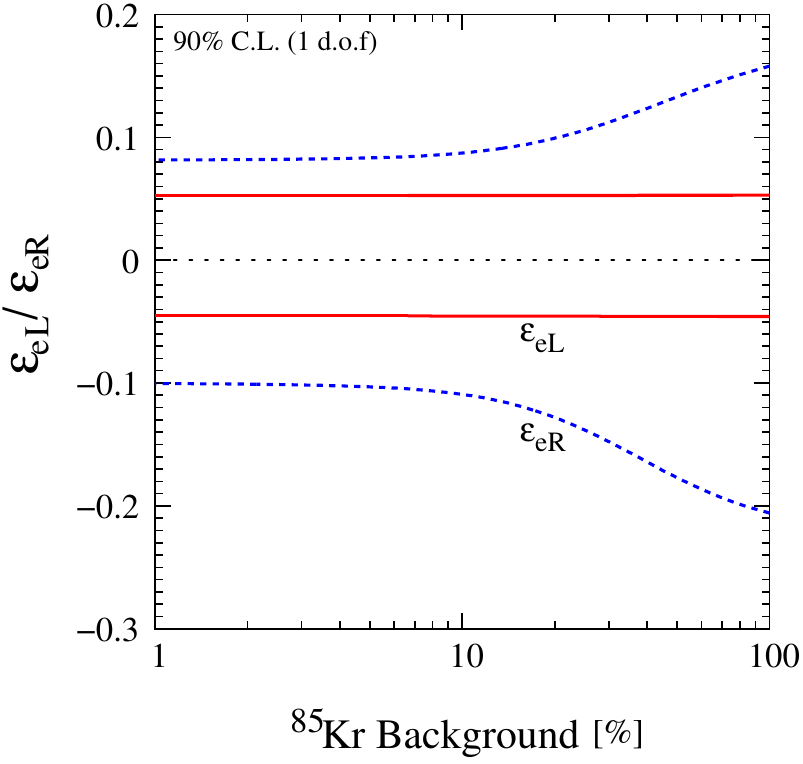}
\includegraphics[width=0.325\textwidth]{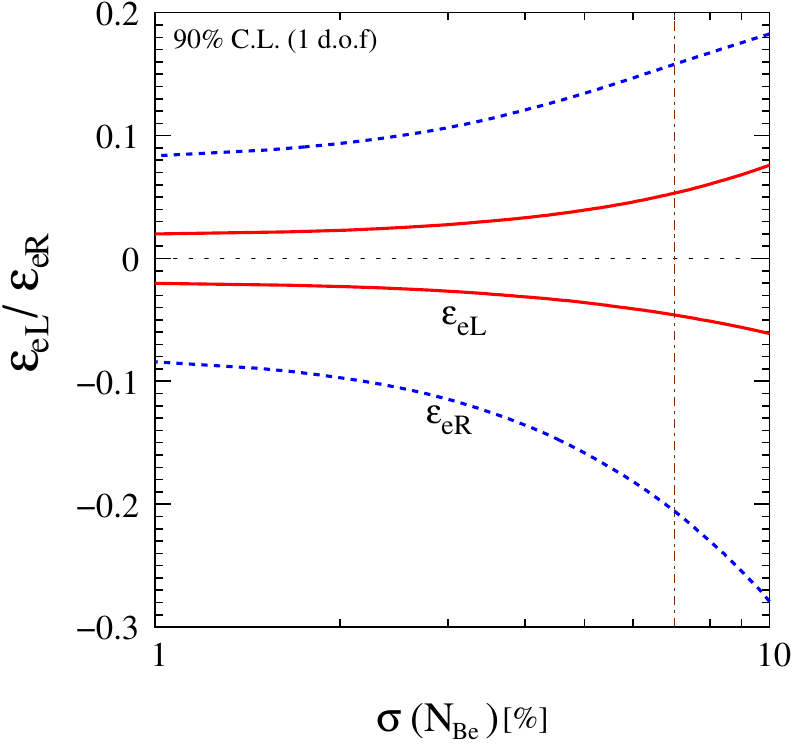}
\caption{\label{fig:eL-eR-factors-vary} 
One NSI parameter at a time limits on $\varepsilon_{eL}$ and $\varepsilon_{eR}$ 
at 90\% C.L. (1 d.o.f) as a function of total fiducial exposure (left panel), total amount of $\Kr$
background (middle panel), and $1\,\sigma$ uncertainty on $\Be$ signal rate (right panel).}
\end{figure}

\begin{figure}[b]
\includegraphics[width=0.325\textwidth]{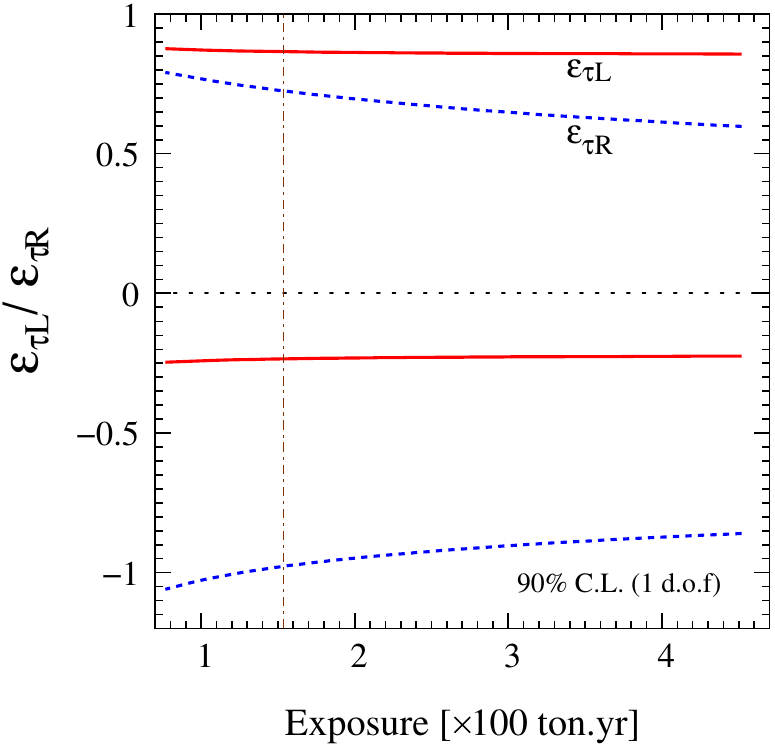}
\includegraphics[width=0.325\textwidth]{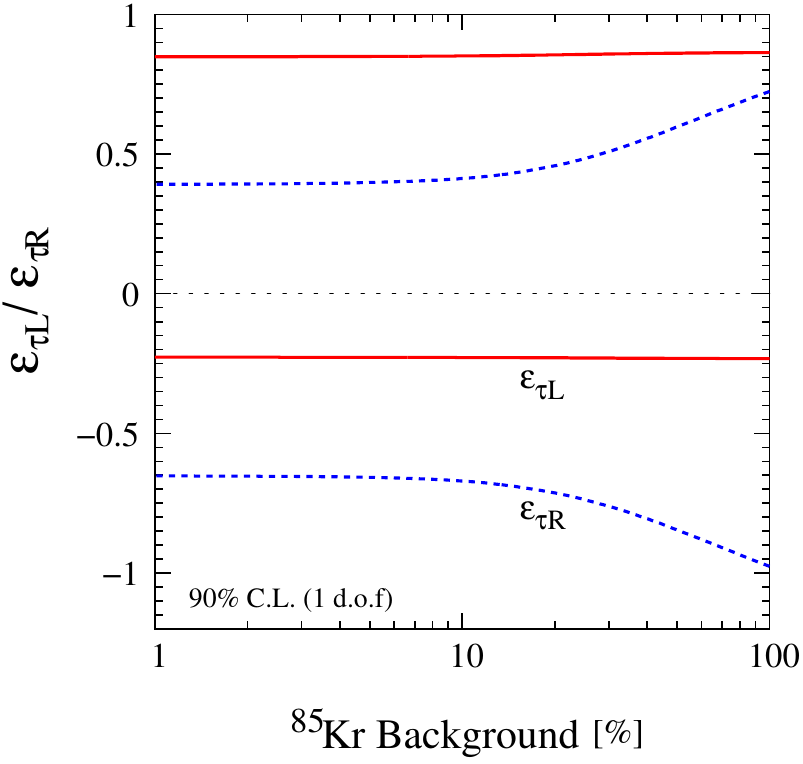}
\includegraphics[width=0.325\textwidth]{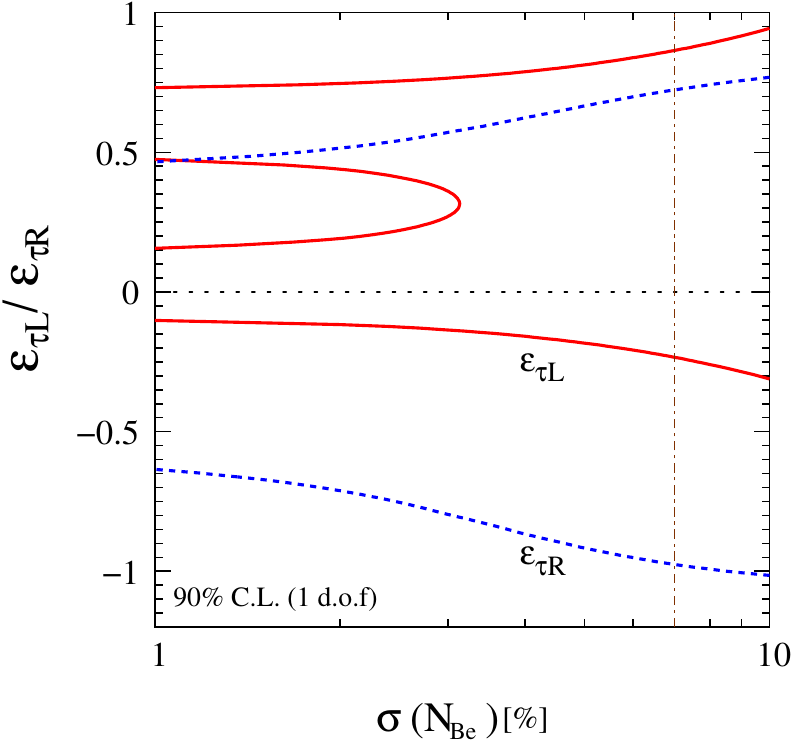}
\caption{\label{fig:tauL-tauR-factors-vary} 
One NSI parameter at a time limits on $\varepsilon_{\tau L}$ and $\varepsilon_{\tau R}$ 
at 90\% C.L. (1 d.o.f) as a function of total fiducial exposure (left panel), total amount of $\Kr$
background (middle panel), and $1\,\sigma$ uncertainty on $\Be$ signal rate (right panel).} 
\end{figure}

Let us quantify the expected improvements on the NSI bounds
due to improved statistics and the reduction of the $\Kr$ background.
In the leftmost panel of Fig.~\ref{fig:eL-eR-factors-vary}, we show how the bounds on the electron NSI parameters
$\varepsilon_{eL}$ and $\varepsilon_{eR}$ will be affected by an increase in the total fiducial exposure while
keeping all other assumptions the same. 
The vertical dot-dashed line indicates the current fiducial exposure of $153.6\;\mathrm{ton\!\cdot\!years}$.  
The leftmost panel of Fig.~\ref{fig:tauL-tauR-factors-vary} shows the same for the tau NSI parameters
$\varepsilon_{\tau L}$ and $\varepsilon_{\tau R}$.
We can see from these panels that increased statistics will not improve the limits on the left-handed coupling at all,
while the improvements in the right-handed couplings are quite modest, indicating that the current
uncertainties in the NSI parameters are mostly due to backgrounds and systematic uncertainties.

%%%%%%%%%%%%%%%%%%%%%%%%%%%%%%%%%%%%%%%%%%%%%%%%%%%%%%%%%%%%%%%%%%%%%%%%%%%%%%
%%%%%%%%%%%%%%%%%%%%%%%%%%%%%%%%%%%%%%%%%%%%%%%%%%%%%%%%%%%%%%%%%%%%%%%%%%%%%%

%%%%%%%%%%%%%%%%%%%%%%%%%%%%%%%%%%%%%%%%%%%%%%%%%%%%%%%%%%%%%%%%%%%%%%%%%%%%%%
%%%%%%%%%%%%%%%%%%%%%%%%%%%%%%%%%%%%%%%%%%%%%%%%%%%%%%%%%%%%%%%%%%%%%%%%%%%%%%

In the center panel of Fig.~\ref{fig:eL-eR-factors-vary}, we show the impact of a reduction in the $\Kr$ background 
on the bounds on $\varepsilon_{eL}$ and $\varepsilon_{eR}$, with a fixed fiducial exposure of $153.6\;\mathrm{ton\!\cdot\!years}$.
The reduction is expressed in percentages compared to Eq.~(\ref{KrDirect}).
The center panel of Fig.~\ref{fig:tauL-tauR-factors-vary} shows the same for $\varepsilon_{\tau L}$ and $\varepsilon_{\tau R}$.
As discussed in section~\ref{sec:event-nsi}, we expect the bounds on the left-handed couplings,
which change the $\Be$ signal normalization, to be little affected 
since the $\Kr$ background mostly changes the slope of the spectrum, and indeed
the figures confirm this expectation.
The bounds on the right-handed couplings, on the other hand, can be tightened.
If the $\Kr$ background is reduced down to 10\%, 
the bounds on $\varepsilon_{e R}$ will shrink by a factor of $\sim\!2$,
while those on $\varepsilon_{\tau R}$ will strink by a factor of $\sim\!1.5$.

In the rightmost panels of
Figs.~\ref{fig:eL-eR-factors-vary} and \ref{fig:tauL-tauR-factors-vary}, 
we show the impact of a reduction in the uncertainty of the $\Be$ signal normalization 
on the bounds on $\varepsilon_{eL}$ and $\varepsilon_{eR}$, and $\varepsilon_{\tau L}$ and $\varepsilon_{\tau R}$, respectively.  
All other assumptions, including the fiducial exposure of $153.6\;\mathrm{ton\!\cdot\!years}$, are kept the same.  
The vertical dot-dashed lines show the current $1\sigma$ uncertainty of $\pm 7\%$. 
If we can reduce this uncertainty from 7\% to, say, 3\% then for $\varepsilon_{eL}$ and $\varepsilon_{eR}$, the limits can be 
improved roughly by a factor of 1.5. For $\varepsilon_{\tau L}$, we find two disjoint regions if we can go below 3\%.   
Such an improvement must first follow the resolution of the solar metalicity problem we
alluded to in section~\ref{sec:Solar_Model}, 
and further improvements in nuclear cross section measurements.

%%%%%%%%%%%%%%%%%%%%%%%%%%%%%%%%%%%%%%%%%%%%%%%%%%%%%%%%%%%%%%%%%%%%%%%%%%%%%%
\subsection{Constraints in the ($\varepsilon_{eL}$-$\varepsilon_{eR}$) 
and ($\varepsilon_{\tau L}$-$\varepsilon_{\tau R}$) planes}
\label{sec:two-parameter}

%%%%%%%%%%%%%%%%%%%%%%%%%%%%%%%%%%%%%%%%%%%%%%%%%%%%%%%%%%%%%%%%%%%%%%%%%%%%%%
\begin{figure}[t]
\includegraphics[width=0.5\textwidth]{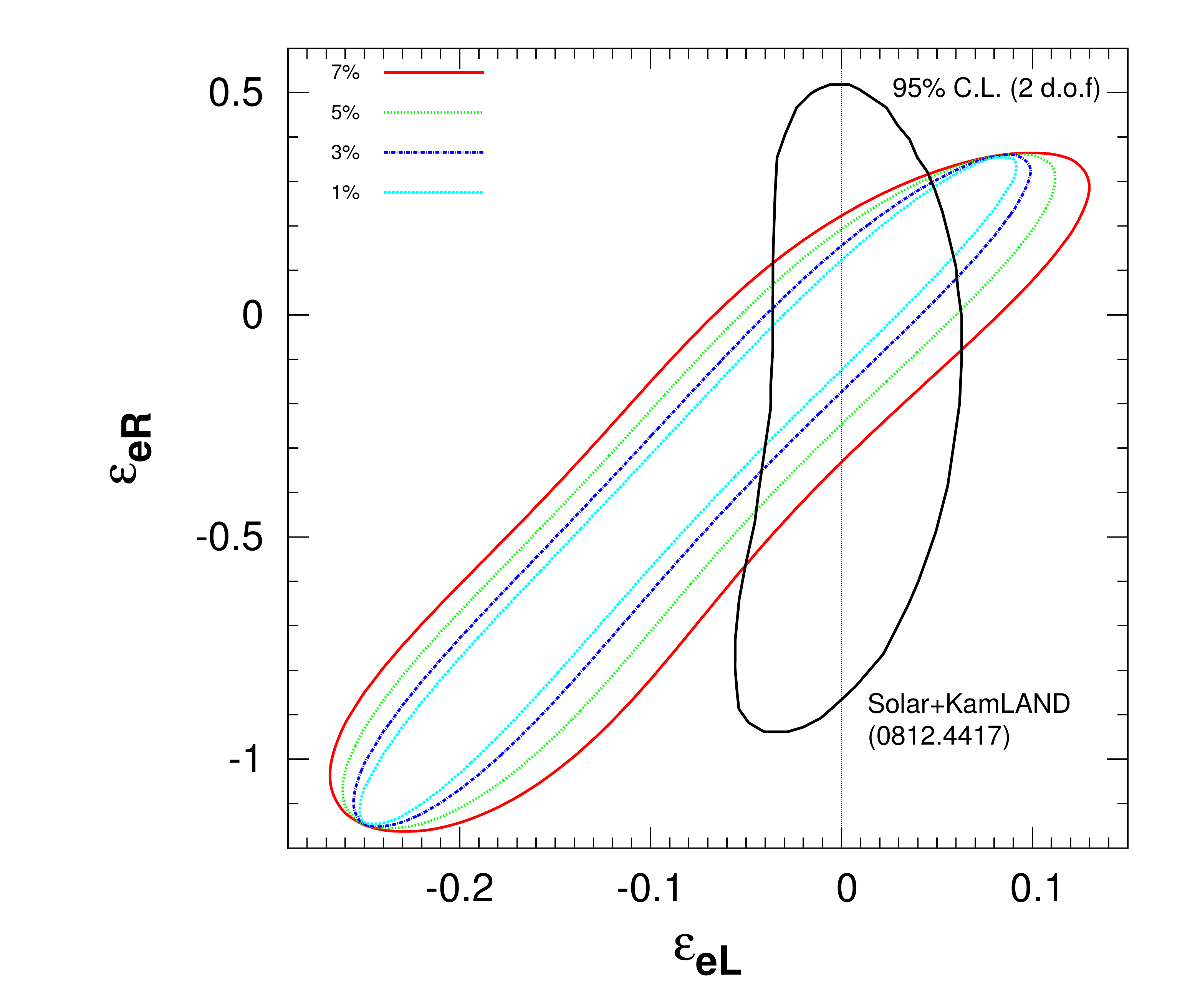}
\includegraphics[width=0.5\textwidth]{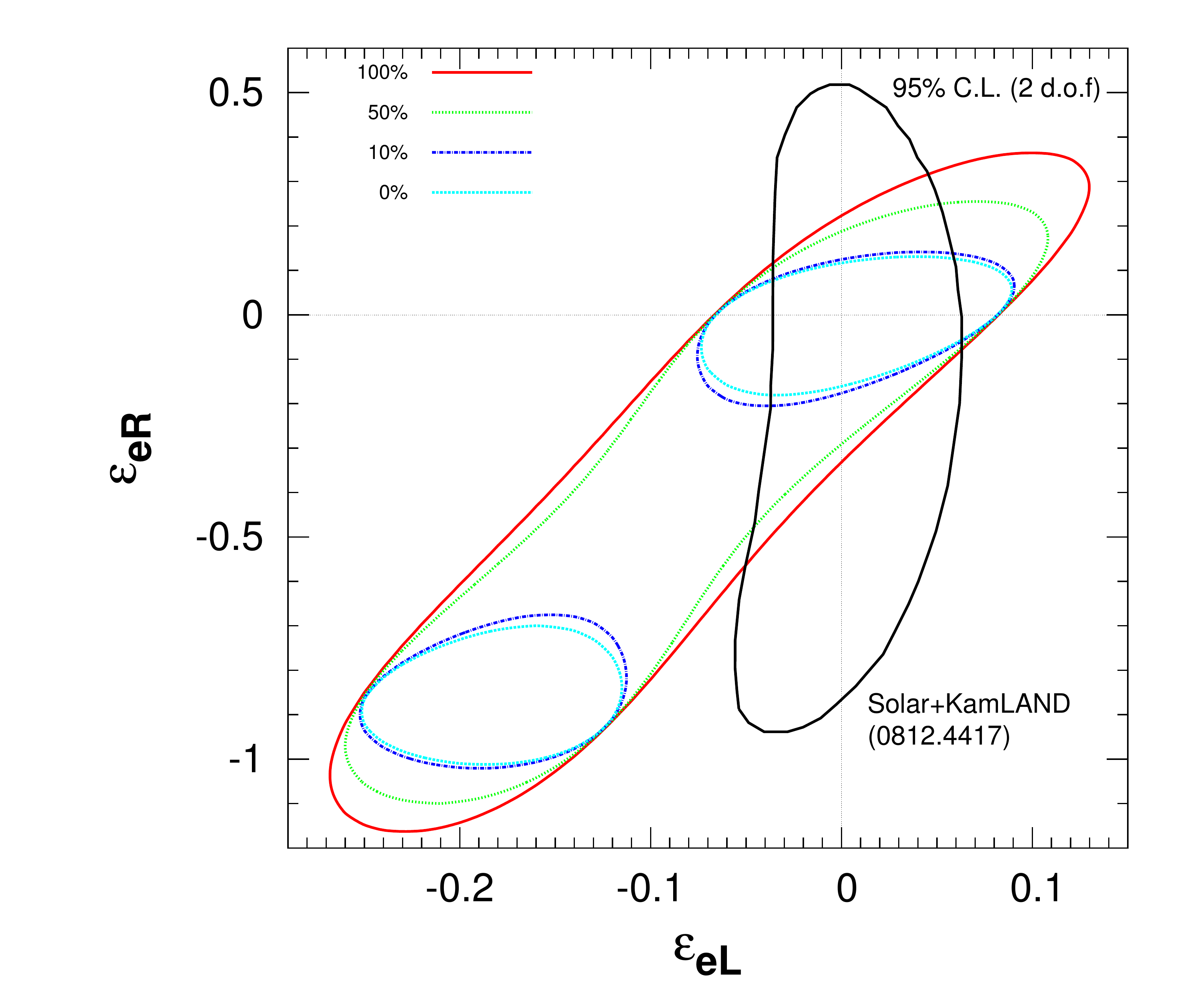}
\caption{\label{fig:eL-eR} 
Allowed regions in the $\varepsilon_{eL}$-$\varepsilon_{eR}$ plane at 95\% C.L. 
(2 d.o.f, $\Delta\chi^2 = 5.99$). 
In this analysis, $\varepsilon_{\tau L}$ and $\varepsilon_{\tau R}$ are fixed to zero,
while the normalization parameters and $s_{23}^2$ are marginalized.
The solid red curves indicate the current bounds, while the green, blue, and cyan curves
indicate what the bounds would be with reduced uncertainty in the $\Be$ signal normalization (left-panel),
and reduced $\Kr$ background (right-panel).
The area outside each contour is excluded.
The bound from Ref.~\cite{Bolanos:2008km} is plotted in black for comparison.
}
\end{figure}
%%%%%%%%%%%%%%%%%%%%%%%%%%%%%%%%%%%%%%%%%%%%%%%%%%%%%%%%%%%%%%%%%%%%%%%%%%%%%%

\begin{figure}[t]
\includegraphics[width=0.5\textwidth]{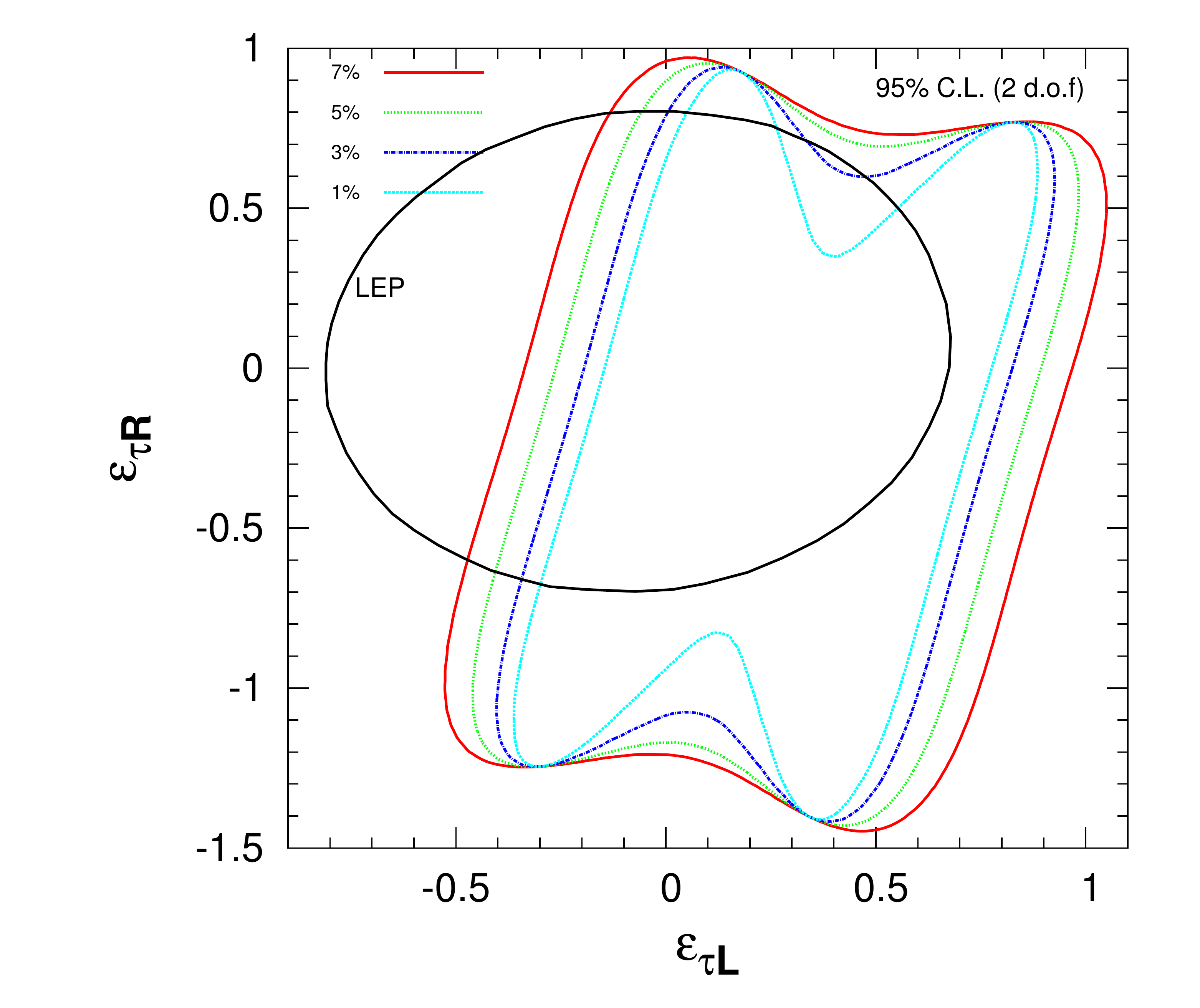}
\includegraphics[width=0.5\textwidth]{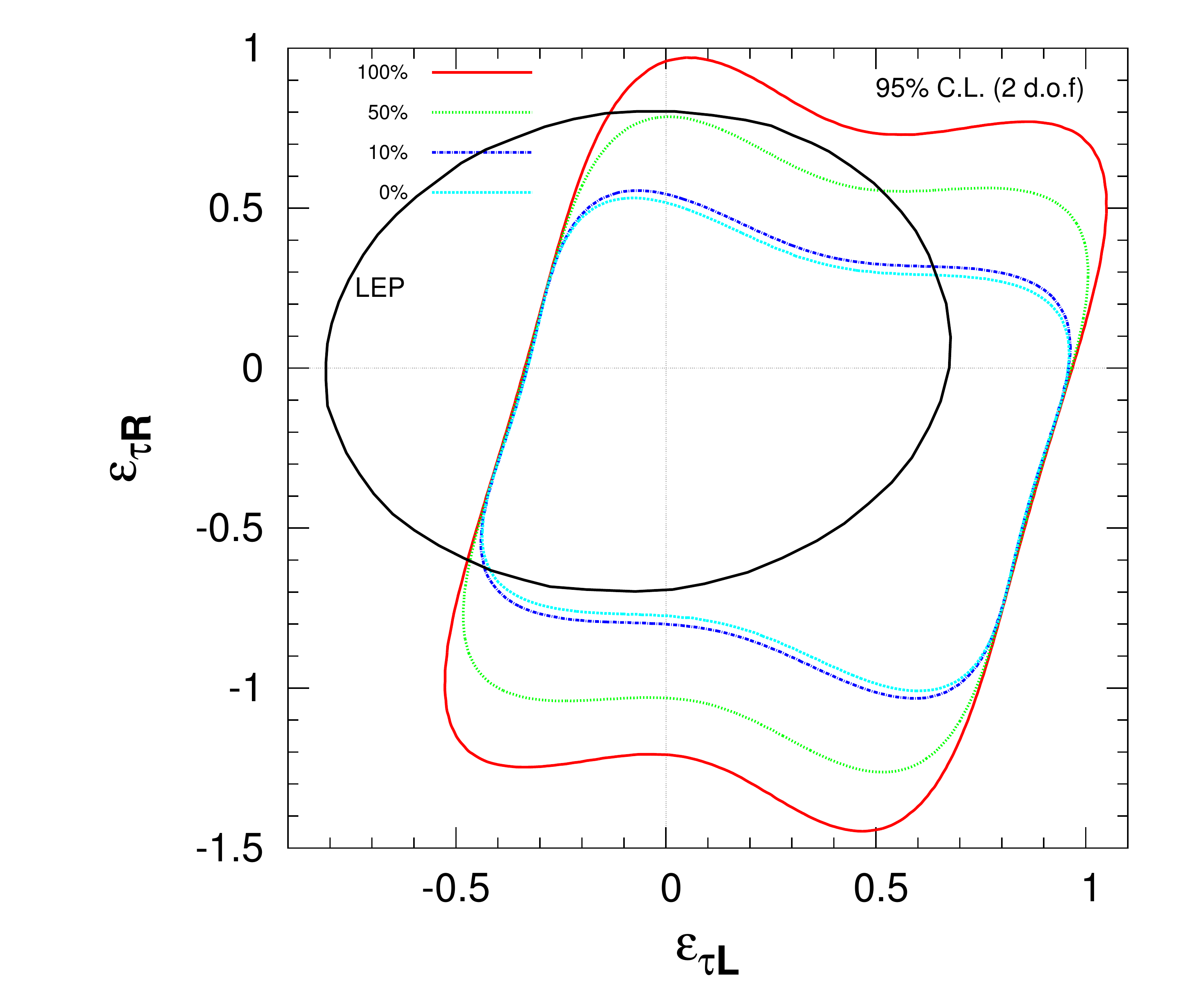}
\caption{\label{fig:tauL-tauR} 
Allowed regions in the $\varepsilon_{\tau L}$-$\varepsilon_{\tau R}$ plane at 95\% C.L. (2 d.o.f, $\Delta\chi^2 = 5.99$). 
In this analysis, $\varepsilon_{e L}$ and $\varepsilon_{e R}$ are fixed to zero,
while the normalization parameters are marginalized.
The solid red curves indicate the current bounds, while the green, blue, and cyan curves
indicate what the bounds would be with reduced uncertainty in the $\Be$ signal normalization (left-panel),
and reduced $\Kr$ background (right-panel).
The area outside each contour is excluded.
The bound from Ref.~\cite{Barranco:2007ej} is plotted in black for comparison.}
\end{figure}

Let us now turn to 2D constraints in the left-right coupling plane of each flavor.
In Fig.~\ref{fig:eL-eR}, we show the allowed regions in the ($\varepsilon_{eL}$-$\varepsilon_{eR}$) plane at  
95\% C.L. (2 d.o.f, $\Delta\chi^2 = 5.99$), obtained with the $\tau$ NSI parameters set to zero,  
$\varepsilon_{\tau L}=\varepsilon_{\tau R}=0$, and the normalization parameters and $s_{23}^2$ marginalized.
On both panels, the solid red curves indicate the current bounds with $153.6\;\mathrm{ton\!\cdot\!year}$ fiducial exposure,
$7\%$ uncertainty in the normalization of the $\Be$~signal, and with the full $\Kr$~background. 
On the left-hand panel, three more curves indicate the bounds assuming three different uncertainty levels 
(see legends in the figure) in the $\Be$~signal normalization. 
On the right-hand panel, three more curves indicate the bounds with three different assumptions 
(see legends in the figure) on the amount of $\Kr$~background. 
For comparison, we plot the combined solar+KamLAND bound obtained in Ref.~\cite{Bolanos:2008km}
in black.
In Fig.~\ref{fig:tauL-tauR}, we show the same for the ($\varepsilon_{\tau L}$-$\varepsilon_{\tau R}$) plane.
For comparison, the bound based on the LEP `neutrino counting' data \cite{Barranco:2007ej}
is plotted in black.

Here, we find that the Borexino bounds are comparable in size to those from Ref.~\cite{Bolanos:2008km}
and \cite{Barranco:2007ej}, but occupy slightly different regions in the 2D parameter space.
Thus, the overlap region in 2D is smaller than either Borexino, or the reference bounds alone.
While the expected improvements to the bounds due to reduced normalization uncertainty and
$\Kr$ background are modest for the $\tau$-parameters, they can be considerable for the 
electron parameters, particularly if the $\Kr$ background can be reduced to $10\%$
of previous levels.

%%%%%%%%%%%%%%%%%%%%%%%%%%%%%%%%%%%%%%%%%%%%%%%%%%%%%%%%%%%%%%%%%%%%%%%%%%%%%%
%%%%%%%%%%%%%%%%%%%%%%%%%%%%%%%%%%%%%%%%%%%%%%%%%%%%%%%%%%%%%%%%%%%%%%%%%%%%%%

%%%%%%%%%%%%%%%%%%%%%%%%%%%%%%%%%%%%%%%%%%%%%%%%%%%%%%%%%%%%%%%%%%%%%%%%%%%%%%
%%%%%%%%%%%%%%%%%%%%%%%%%%%%%%%%%%%%%%%%%%%%%%%%%%%%%%%%%%%%%%%%%%%%%%%%%%%%%%
\section{Summary}
\label{sec:summary}
%%%%%%%%%%%%%%%%%%%%%%

We have used the $153.6\;\mathrm{ton\!\cdot\!year}$ fiducial exposure data from
Borexino to place bounds on the flavor-diagonal NSI parameters
$\varepsilon_{e L}$, $\varepsilon_{e R}$, $\varepsilon_{\tau L}$, and
$\varepsilon_{\tau R}$ taking into account the $\pm 7\%$
uncertainty in the $\Be$ solar neutrino flux, $\Phi_{\Be}^{0.862}$,
and the backgrounds from $\Kr$ and $\Bi{210}$ $\beta$-decay.
The uncertainty in $\sin^2\theta_{23}$ was assumed to be $\pm 11\%$ around
the reference value of $0.5$.
The resulting one NSI parameter at a time bounds are listed in 
Table~\ref{tab:limits}. 
2D bounds in the ($\varepsilon_{eL}$-$\varepsilon_{eR}$) 
and ($\varepsilon_{\tau L}$-$\varepsilon_{\tau R}$) planes are shown in
figures~\ref{fig:eL-eR} and \ref{fig:tauL-tauR}.
They are comparable to existing bounds in Ref.~\cite{Bolanos:2008km} and \cite{Barranco:2007ej}.

Further improvements in the bounds would require reductions in the
$\Kr$ background, which is already underway in Borexino Phase~II, and in the uncertainty of
the $\Be$ solar neutrino flux. 
The latter may be achieved by the resolution of the solar metalicity problem,
and improvements in the relevant cross section measurements by future
experiments such as DIANA \cite{DIANA:2009}.

%%%%%%%%%%%%%%%%%%%%%%%%%%%%%%%%%%%%%%%%%%%%%%%%%%%%%%%%%%%%%%%%%%%%%%%%%%%%%%
%%%%%%%%%%%%%%%%%%%%%%%%%%%%%%%%%%%%%%%%%%%%%%%%%%%%%%%%%%%%%%%%%%%%%%%%%%%%%%
\acknowledgments
%%%%%%%%%%%%%%%%%%

We would like to thank Carlos Pe\~na-Garay, Patrick Huber, Jonathan Link,
and Bruce Vogelaar for helpful discussions.
We would also like to thank Yee Kao for his contributions to the early stages of this work.
TT is grateful for the hospitality of the Instituto de F\'isica Corpuscular
(IFIC) at the University of Valencia, Spain, 
where a large portion of this work was carried out during his visit in June 2012.
SKA acknowledges the support from the European Union under 
the European Commission FP7 Research Infrastructure Design Studies 
EUROnu (Grant Agreement No.~212372 FP7-INFRA-2007-1), 
LAGUNA (Grant Agreement No. 212343 FP7-INFRA-2007-1), 
and the project Consolider-Ingenio CUP. 
FL is supported by the Universit\`{a} degli Studi dell'Aquila, 
Dipartimento di Fisica, and by a PhD scholarship from the
Laboratori Nazionali del Gran Sasso (INFN-LNGS). 
TT is supported in part by the U.S. Department of Energy, grant DE-FG05-92ER40677, task A.

\bigskip

%\textit{This work is dedicated to the memory of Raju Raghavan,
%who was the father of the Borexino experiment, and
%the initiator and driving force behind this work.}

%%%%%%%%%%%%%%%%%%%%%%%%%%%%%%%%%%%%%%%%%%%%%%%%%%%%%%%%%%%%%%%%%%%%%%%%%%%%%%
%%%%%%%%%%%%%%%%%%%%%%%%%%%%%%%%%%%%%%%%%%%%%%%%%%%%%%%%%%%%%%%%%%%%%%%%%%%%%%
%\newpage
\bibliographystyle{JHEP}
\bibliography{nsi-references}

\end{document}